\documentclass{nature} 
\usepackage{graphicx} 
\usepackage{outlines}
\usepackage{xcolor}
\usepackage[colorlinks=false]{hyperref}
\usepackage{algorithm}
\usepackage{algorithmicx}
\usepackage{lineno}
\usepackage{amsmath}
\usepackage{siunitx}
\usepackage[font={small,singlespacing},labelfont=bf]{caption}
\usepackage{titling}

\title{Design framework for programmable three-dimensional woven metamaterials}
\author{Molly Carton$^{1,2}$, James Utama Surjadi$^{1}$, Bastien F. G. Aymon$^{1}$, Carlos M.~Portela$^{1\ast}$}
\begin{document}
\maketitle

\begin{affiliations}
\footnotesize
 \item Department of Mechanical Engineering, Massachusetts Institute of Technology, Cambridge, MA 02139, USA
 \item Department   of   Mechanical   Engineering,   University of Maryland, College Park, MD, USA
\end{affiliations}
\vspace{10pt}
\spacing{1.0}

\begin{abstract}

Mechanical metamaterials have continued to offer unprecedented tunability in mechanical properties, but most designs to date have prioritized attaining high stiffness and strength while sacrificing deformability. The emergence of woven lattices---three-dimensional networks of entangled fibers---has introduced a pathway to the largely overlooked compliant and stretchable regime of metamaterials. However, the design and implementation of these complex architectures has remained a primarily manual process, restricting identification and validation of their full achievable design and property space. Here, we present a geometric design framework that encodes woven topology using a graph structure, enabling the creation of woven lattices with tunable architectures, functional gradients, and arbitrary heterogeneity. Through use of microscale \emph{in situ} tension experiments and computational mechanics models, we reveal highly tunable anisotropic stiffness (varying by over an order of magnitude) and extreme stretchability (up to a stretch of four) within the design space produced by the framework. Moreover, we demonstrate the ability of woven metamaterials to exhibit programmable failure patterns by leveraging tunability in the design process. This framework provides a design and modeling toolbox to access this previously unattainable high-compliance regime of mechanical metamaterials, enabling  programmable large-deformation, nonlinear responses.

\end{abstract}
\newpage

\subsection{Introduction}
\hfill\\
\indent The advancement of additive manufacturing technologies has solved many fabrication challenges, enabling the ability to print complex internal structures across length scales or the possibility to print highly compliant and slender three-dimensional (3D) structures. These new possibilities have driven interest in the development of increasingly complex 3D architected materials---or mechanical metamaterials---as evidenced by this field's accelerated progress in recent years\cite{bauer2017nanolattices, surjadi2019mechanical, greer2019three, schwaiger2019extreme,surjadi_enabling_2025}. Architected materials have been demonstrated to exhibit a variety of complex and unconventional behaviors, including high strength and stiffness at low densities\cite{zheng_ultralight_2014, meza2014strong, bauer_approaching_2016}, energy absorption and impact resistance\cite{shan2015multistable, portela2020extreme, portela2021supersonic, surjadi2025exploiting}, tunable Poisson's ratio\cite{teng2022simple, farzaneh2022sequential}, negative thermal expansion\cite{yang20203d}, acoustic waveguiding\cite{bayat2018wave}, and vibration absorption\cite{matlack2016composite}. While thousands of microstructures have been proposed over the last decade, most design paradigms of architected materials can be broadly classified into three main categories, namely truss-\cite{zheng2014ultralight, meza2014strong, bauer_approaching_2016}, shell-\cite{al2018microarchitected, portela2020extreme}, and plate-based\cite{berger_mechanical_2017, crook2020plate} architectures. These categories emerged naturally in the search for record-breaking stiff and strong materials, where optimized mechanical properties for a given material density were sought. In this race to achieve ultra‑high stiffness and exceptional strength‑to‑weight ratios, the equally important pursuit of highly compliant and deformable architectures has remained largely unexplored.

Designing compliant architected materials presents opportunities to populate uncharted areas for this materials system, in a domain where they can offer promising solutions for soft-matter applications such as flexible sensors\cite{luo2023technology} and bioscaffolds\cite{wang2022bioinspired}. To achieve enhanced compliance and deformability, variations of conventional design categories---i.e., variations of truss-, shell-, or plate-based architectures---with functional gradients\cite{rodrigo2021crushing, rodrigo2023mechanical, noronha2023additively} or heterogeneity\cite{pham2019damage, xiao20213d, liu2024spatially, chen2024heterostructured} are often proposed. For example, density-graded architectures have been used to reduce the initial stiffness of truss-based lattices while enabling continuous increase in engineering stress, resulting in higher peak stress and dissipated energy density\cite{rodrigo2021crushing}. Functional gradients have also been utilized to inhibit the formation of shear bands by inducing layer-by-layer deformation, which prevents catastrophic failure\cite{noronha2023additively}. In addition, heterogenous lattices that combine multiple types of unit cells in the same sample have been implemented to tailor the deformation or failure of lattices, allowing for pre-determined failure paths by designing regions with different compliance\cite{pham2019damage, xiao20213d, liu2024spatially, chen2024heterostructured}. Nevertheless, drastically improving the compliance and deformability of architected materials solely by varying the topological arrangement of classical architectures is challenging, especially under tension, owing to the intrinsically stiff nature of these designs.

Recently, new classes of architected materials, such as 3D helical\cite{yan2020soft} and woven lattices\cite{moestopo2020pushing, moestopo2023knots,surjadi_double-network-inspired_2025}, have been proposed to achieve extreme combinations of compliance and deformability. These designs leverage the structural flexibility of coiled helical fibers, as opposed to the straight beams found in truss-based lattices, to reduce stiffness while enhancing stretchability through fiber straightening. In particular, woven lattices differ from conventional truss- or plate-based lattices with monolithic junctions at joints/nodes by instead forming entanglements of fibers with constant curvature at junctions. As a result, stress concentration is reduced at these entangled junctions during deformation, thus enabling large deformations (with stretch $\lambda$ values exceeding 2) even with stiff and brittle constituent materials\cite{moestopo2020pushing, moestopo2023knots}. Moreover, frictional sliding\cite{karapiperisFrictionBeam2024} and the evolution of entanglements between woven fibers during deformation provide additional energy dissipation mechanisms\cite{ryan2015damping, salari2018damping, garland2020coulombic, li2022harnessing, moestopo2020pushing, moestopo2023knots,surjadi_double-network-inspired_2025}, resulting in high toughness. However, current studies on woven lattices have been limited to periodic tessellations of two main architectures (octahedron and diamond)\cite{moestopo2020pushing} and their simpler variations\cite{moestopo2023knots}, restricting the full range of achievable properties in this materials system. The lack of variation and tunability in woven lattice designs can be attributed to the difficulty of parametrizing their complex hierarchical nature, consequently leading to the absence of robust design frameworks capable of efficiently generating woven lattices with arbitrary 3D topologies---without the need for manual design in computer-aided design software. 

Here, we introduce a design and modeling framework for 3D woven metamaterials---validated by experiments on a range of newly achievable designs---that vastly expands the usable design space of these promising materials. We introduce a general route for converting arbitrary beam lattices into 3D woven lattices using their graph representation\cite{zheng_unifying_2023}. As a demonstration, we introduce three previously unachievable woven topologies. By defining the design space of a given topology with two strut-level parameters, i.e., the effective strut radius and the number of fiber revolutions within, we allow facile design of arbitrary functionally graded and spatially patterned lattices without loss of geometric compatibility between differing unit cells. Using computational mechanics models directly coupled to the design-generation framework, we establish the broad range of elastic behaviors achievable by solely varying unit cell and strut parameters. Most importantly, we present a reduced-order beam-element representation for nonlinear, large deformation which allows rapid exploration of the full nonlinear design space for large tessellations of this metamaterial class, which is validated against \emph{in situ} tension experiments on microscale woven lattices. Specifically, we investigate their characteristic deformation and the effects of fiber topology (i.e., the path of a given continuous fiber in a 3D domain) on the mechanical responses of the metamaterials, while identifying fiber straightening and entanglement evolution as key deformation mechanisms. This validated framework---provided herein as a freely available software design tool---enables broader implementation of woven metamaterials as a newly established compliant-material class within the mechanical metamaterials toolbox. 

\subsection{Results}
\begin{figure}[H] 
\centering
\includegraphics[width=.99\textwidth]{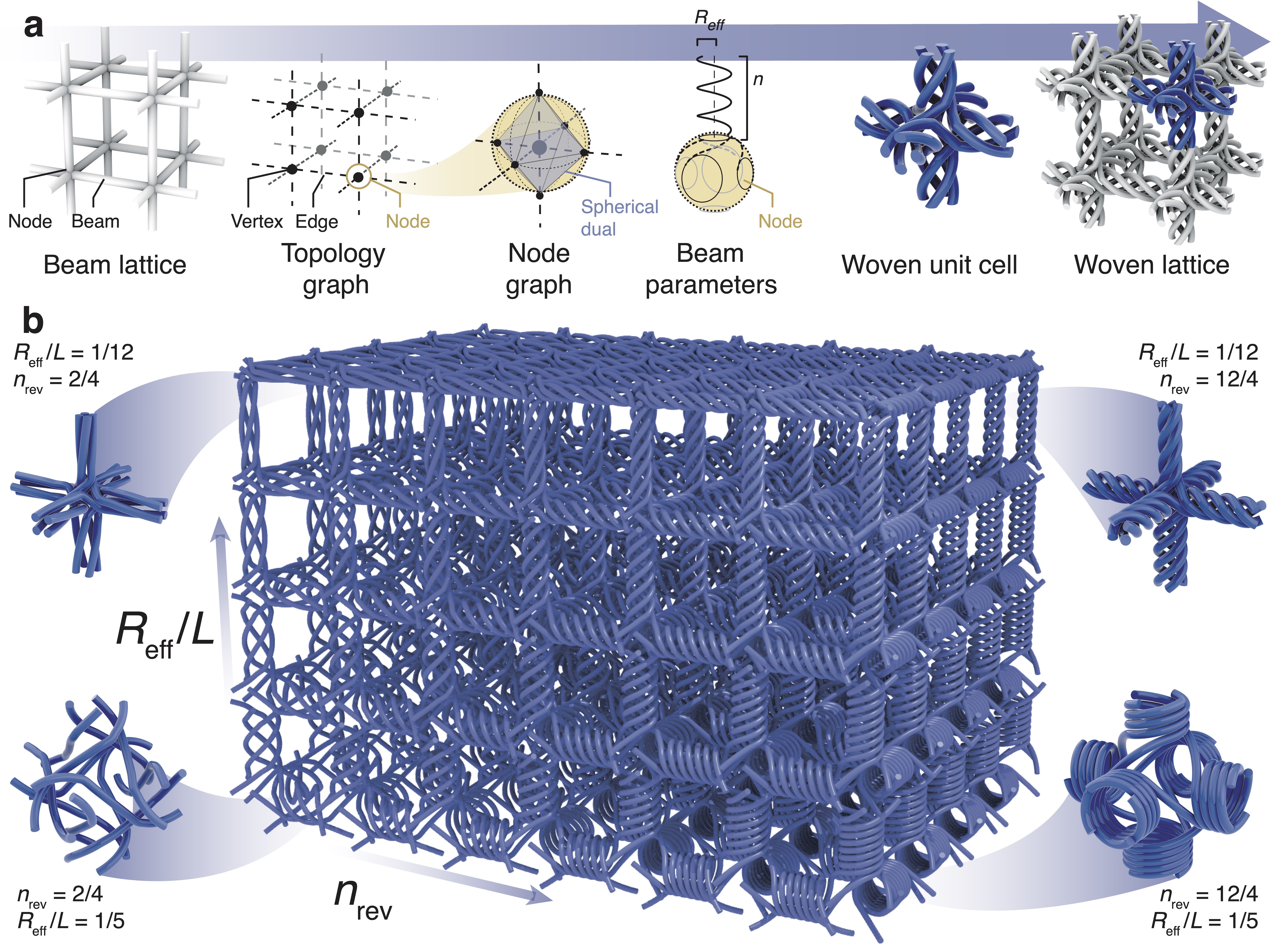}
\caption{\textbf{Design and realization of parametric 3D woven lattices}. {\bf a,} Schematic of the framework's translation between an example cubic beam-and-node lattice and the resulting woven cubic lattice, with intervening graph and woven-beam-parameter representations. {\bf b,} CAD render of a functionally graded cubic lattice produced by the framework, possessing functional grading in: (i) effective radius $R_{\text{eff}}$ in the vertical direction, from $\frac{1}{5}$ to  $\frac{1}{12}$ of the unit cell side length $L$; and (ii) number of turns $n_{\text{rev}}$ of an individual fiber along a specific beam, from $\frac{2}{n}$ to $\frac{12}{n}$ revolutions per beam, where---in the cubic case---$n=4$ represents the connectivity of the node graph (and hence the number of fibers comprising each beam).}
\label{fig:overview}
\end{figure}

\hfill\\ \\ 
\noindent\emph{A Rational Design Framework for Woven Metamaterials}
\hfill\\
\indent To fully determine the design space of 3D woven metamaterials and establish relations between design parameters and resulting macroscopic properties, we developed a systematic framework for their rational design (Fig.~\ref{fig:overview}a). In short, the generation of woven lattices begins with a classification based on the parent monolithic-beam lattice, translated to a spatial-graph-based representation of the topology (i.e., the topology graph) as vertices and edges. The vertices in the topology graph are expanded into spherical node graphs, which dictate how fibers in a woven beam connect to adjacent woven beams. Each edge in the topology graph that meets the spherical node is then replaced with an effective woven beam---a construct formed by helical fiber bundles---characterized by an effective diameter and the revolutions of its constituent fibers. To connect effective woven beams, including those with dissimilar geometric parameters, the woven nodes are engineered using a chiral twist that seamlessly connects each effective beam to its topological neighbors, ensuring continuity across all fiber strands. Thus, this simple graph-based approach facilitates the design of large tessellations of woven unit cells, by effectively capturing the topology of the woven lattice, and embedding the complexity of the woven nodes and beams as scalar parameters of the graph's vertices and edges. 

Once the topology graph for an envisioned woven lattice is defined, the more involved step consists of generating the graph representation of its woven nodes. The vertices of these spherical node graphs are determined by the intersection of a woven beam's centerline with the spherical node region. 
Joining all vertices with edges forms the faces of a polyhedron (as shown in Fig.~\ref{fig:overview}a Node) within this spherical region. The resulting nodal graph formed by these vertices and edges (Fig.~S1b) is the spherical dual of the topology graph, which represents connections between spatially neighboring edges\cite{heath_representing_1993}. Connecting neighboring beams accordingly enforces the fundamental criterion that each beam must interweave with all its nearest neighbors, reconstructing the full connectivity of the original beam lattice. Additionally, this representation introduces a structural constraint that the degree (i.e., connectivity) of each nodal graph must remain constant, which guarantees that each beam in the woven lattice is composed of an equal number of intertwined fibers. In other words, a topology graph of $n$-degree produces a woven lattice whose beams are composed of $n$ helical fibers. For the case of a cubic lattice, depicted in Fig.~\ref{fig:overview}a, the spherical dual graph of a cube has the connectivity of an octahedron; each vertex has degree $n=4$ and the resulting woven lattice is thus composed of 4-helices. 

After defining both the topology graph and the nodal graph, the first user-defined parameter of the lattice is the number of turns of a helix $n_\mathrm{rev}$ within a woven beam. This parameter is a discrete, unitless number assigned to each woven beam, which dictates the helix pitch and ultimately how the $n$ fibers of one beam twist and connect to the $n$ fibers in adjacent beams. 
This twist determines the chirality of these helices, which is arbitrary, but remains constant over the entire lattice. 
At this stage, the representation contains the topological information required to define the woven lattice.
The second user-defined parameter of the lattice is the effective radius $R_{\text{eff}}$ of the woven-beam helices, which determines both the shape of the helix and the radius of the node, ensuring a smoothly varying curve path between neighboring helices. This parameter is defined so that the original monolithic-beam lattice is recovered when $R_{\text{eff}}$ goes to zero. At this stage, the lattice is represented as a set of three-dimensional space curves forming the centerlines of fibers, which can then be discretized by sampling points along the centerlines. To obtain a solid geometry, the curves are then swept with a circular profile of a specified fiber radius (details in \emph{Supplementary Information S1} and Fig.~S1).  

This geometric representation therefore preserves the defining features of woven lattices, i.e., an effective beam structure composed of fibers that are highly entangled in all directions at nodes, while allowing for high tunability. While $R_{\text{eff}}$ controls the density and spacing of woven fibers, $n_\mathrm{rev}$ governs fiber entanglement since higher values result in more twisted fibers, while also determining the path by which a fiber traverses the lattice. For example, in the case of a cubic lattice with a specified $n_\mathrm{rev}=3n/4$ for integer $n$, each fiber exits the node along the same cubic face it enters, causing the fiber path to follow the edges of a cubic face (Fig.~S5). This configuration results in a lattice of interlocking square rings, where the integer value of $n_\mathrm{rev}$ dictates how many times each ring interlocks with its neighboring rings. Conversely, other values of constant $n_\mathrm{rev}$ create fiber paths that traverse multiple unit cells. Therefore, since $n_\mathrm{rev}$ determines the global fiber topology, it has a direct effect on the deformation of the woven lattice. A stochastic selection of $n_\mathrm{rev}$ generates a randomly entangled network that retains structured order at the lattice scale.    
The tunability of the described framework thus allows for the generation of spatially varying woven lattices by allowing $n_\mathrm{rev}$ and $R_{\text{eff}}$ to vary as functions of position within the lattice (Figure~\ref{fig:overview}b, \emph{Supplementary Information S2} and Fig.~S2). By spatially modulating $n_\mathrm{rev}$, one can transition from highly entangled regions to more open, loosely connected domains, enabling controlled gradients in mechanical and functional properties. Similarly, varying $R_{\text{eff}}$ allows for localized tuning of effective beam stiffness and node size, accommodating heterogeneous structural requirements. This approach is particularly useful for designing functionally graded materials, where different regions of a structure demand distinct mechanical behaviors. Furthermore, leveraging computational optimization techniques can facilitate the automated design of spatially varying woven lattices tailored to target performance criteria, opening new avenues for advanced metamaterial design and multifunctional applications. Towards facilitating generation of arbitrary woven-unit-cell tessellations, the design framework outlined in this section enabled the creation of an automated design tool provided as Supplementary Software.

\begin{figure}[H] 
\centering 
\includegraphics[width=.9\textwidth]{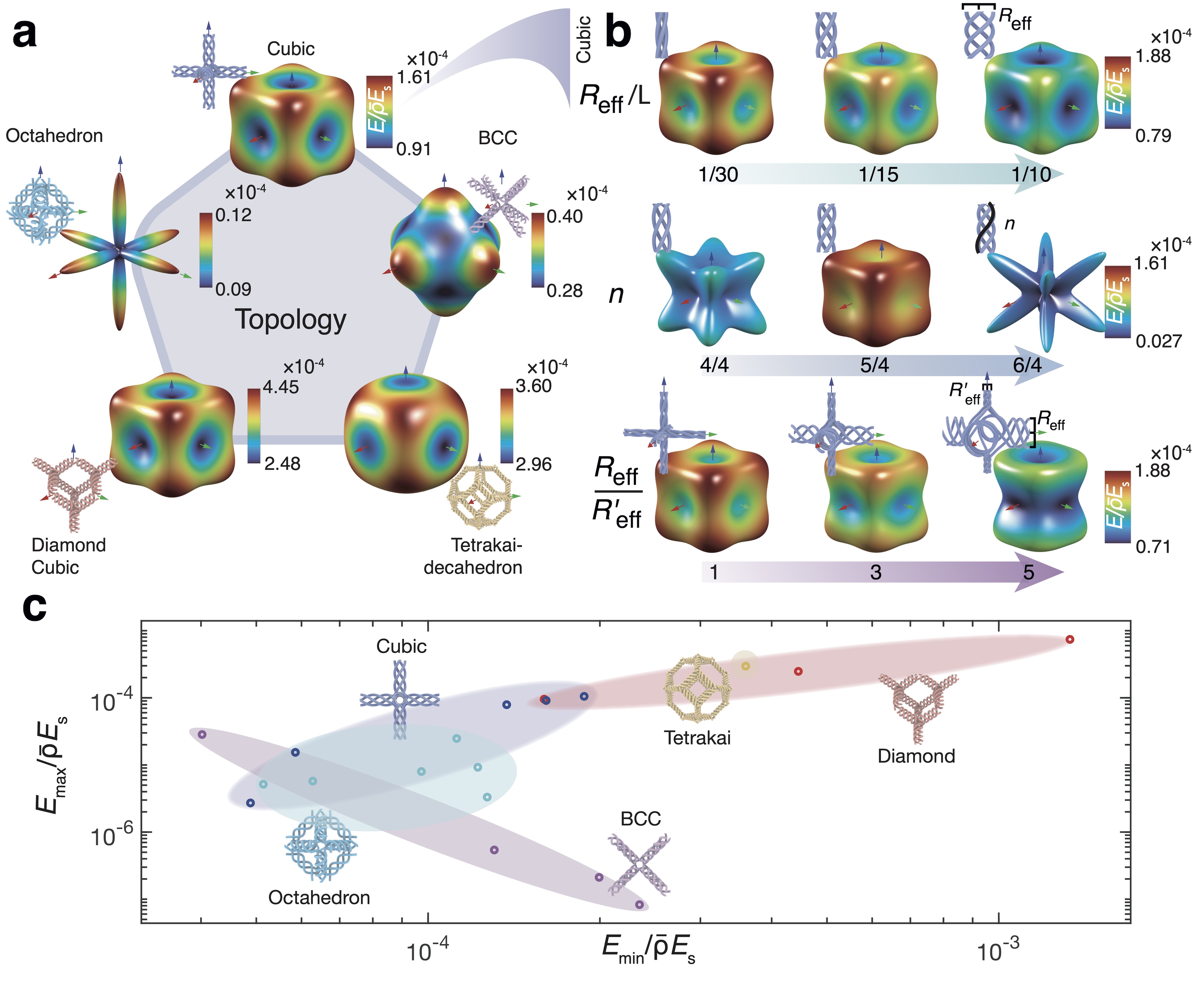}
\caption{\textbf{Linear anisotropy of woven lattices.} 
\textbf{a}, Linear perturbation homogenization analysis comparing elastic surfaces among woven lattice topologies, demonstrating the effect of topology on stiffness and anisotropy. The elastic surfaces represent the normalized maximum specific modulus, corresponding to $E_{\text{max}}$ normalized by the product of the material's relative density (or fill fraction) $\bar{\rho}$ and the contituent material's Young's modulus $E_{\text{s}}$. Clockwise from top, representative unit cells for each demonstrated woven lattice: cubic, $n_{\mathrm{rev}} = 5/4$ and $R_{\mathrm{eff}}/L = 1/15$; body-centered cubic (BCC), $n_{\mathrm{rev}} = 5/3$ and $R_{\mathrm{eff}}/L = 1/15$; tetrakaidecahedron, $n_{\mathrm{rev}} = 4/3$ and $R_{\mathrm{eff}}/L = 1/30$; diamond cubic (diamond), $n_{\mathrm{rev}} = 4/3$ and $R_{\mathrm{eff}}/L = 1/15$; octahedron, $n_{\mathrm{rev}} = 4/3$ and $R_{\mathrm{eff}}/L = 1/15$. All unit cells, fiber radius of $L/60$.
\textbf{b}, Varying parameters $R_{\mathrm{eff}}$ and  $n_{\mathrm{rev}}$ on a cubic unit cell demonstrates the effect of tunable stiffness and anisotropy behavior. Furthermore, introducing dissimilar $R_{\mathrm{eff}}$ values within a single unit cells provides control of anisotropy, as demonstrated in a cubic unit cell with directionally varied $R_{\mathrm{eff}}$ (bottom).
\textbf{c}, Space of normalized maximum specific modulus achieved through variation of $R_{\mathrm{eff}}$ and $n_{\mathrm{rev}}$ for the topologies and parameter space described in this work.} 
\label{fig:linear}
\end{figure}

\noindent\emph{Linear Property Space and Anisotropy}
\hfill\\
\indent The mechanical response of woven lattices is inherently tunable through variations in lattice topology and geometric parameters, enabling a diverse range of anisotropic stiffness behaviors. To evaluate the mechanical tunability of various woven designs, their linear-elastic responses were determined using computational homogenization. By applying periodic boundary conditions at the edges of the unit cell and imposing a series of linearly independent strain inputs, the elasticity tensor for each woven structure was calculated (see Methods). The resulting omnidirectional stiffness, normalized by the Young's modulus of the constituent material, was visualized as elastic surfaces to illustrate the anisotropic characteristics of each architecture (Fig.~\ref{fig:linear}a, Fig.~S3). We identify notable changes in anisotropy for some lattice topologies when converting them from monolithic to woven beams. For instance, lattices with strongly stretching-dominated orientations, such as the cubic and octahedron lattices (both in the [100] direction), undergo drastic anisotropy changes. In their woven forms, both lattices are no longer stretching-dominated in any direction, and this is visible as a distinct change in their elastic surfaces (Fig.~S4). 

In contrast to monolithic lattices, a given woven lattice topology can provide a diverse range of anisotropic responses, through control of geometric parameters such as $R_{\text{eff}}$ and $n_{\text{rev}}$. For instance, changing $R_{\text{eff}}/L$ of a cubic woven lattice, where $L$ is the unit cell size, from $1/30$ to $1/5$ results in a decrease of the normalized maximum specific modulus $E_{\text{max}}/\bar{\rho}E_{\text{s}}$---defined as the maximum directional stiffness $E_{\text{max}}$ normalized by the product of the material's relative density (or fill fraction) $\bar{\rho}$ and the contituent materials' Young's modulus $E_{\text{s}}$---by a factor of more than 2 (Fig.~\ref{fig:linear}b, top). This behavior is similar to that of a classic torsion spring, where an increase in the helix radius results in reduced stiffness. Since $R_{\text{eff}}$ can be tuned continuously, it enables a continuous transition between different stiffness regimes. On the other hand, varying $n_\mathrm{rev}$ alters both stiffness and anisotropy (Fig.~\ref{fig:linear}b, middle), which can be explained by the difference in fiber topology determined by the changing connections between beams (Fig.~S5). To further demonstrate the design flexibility enabled by this framework, we also show that a parameter change can tune anisotropy within an individual unit cell (Fig.~\ref{fig:linear}b, bottom). By selecting dissimilar $R_{\text{eff}}$ values for the beams along one direction and those orthogonal to it (defined by a ratio $R_{\text{eff}}/R'_{\text{eff}}$), the unit cell develops preferentially stiff directions, as demonstrated by the change in shape of the elastic surface. This ability to tune geometric parameters at the sub-unit-cell scale demonstrates localized control over mechanical properties, expanding the property space of each lattice topology.

Plotting the maximum and minimum specific moduli, normalized by relative density $\bar{\rho}$, for a range of lattice topologies further illustrates the tunability of woven lattices in the elastic regime (Fig.~\ref{fig:linear}c). 
This stiffness parameter space reveals a distinct clustering of structures based on their underlying lattice topology, which sets the fundamental mechanical signature of a woven lattice. However, within each topology, moderate changes in $R_{\text{eff}}$, $n_{\mathrm{rev}}$, and $R_{\text{eff}}/R'_{\text{eff}}$ allow for smooth transitions across this property space, unlocking stiffness and anisotropy ranges that approach one order of magnitude. This tunability suggests that woven lattices, unlike their monolithic counterparts, occupy a continuous design space where mechanical performance can be selectively programmed over a range that spans across multiple orders of magnitude.

\begin{figure}[H] 
\centering 
\includegraphics[width=.8\textwidth]{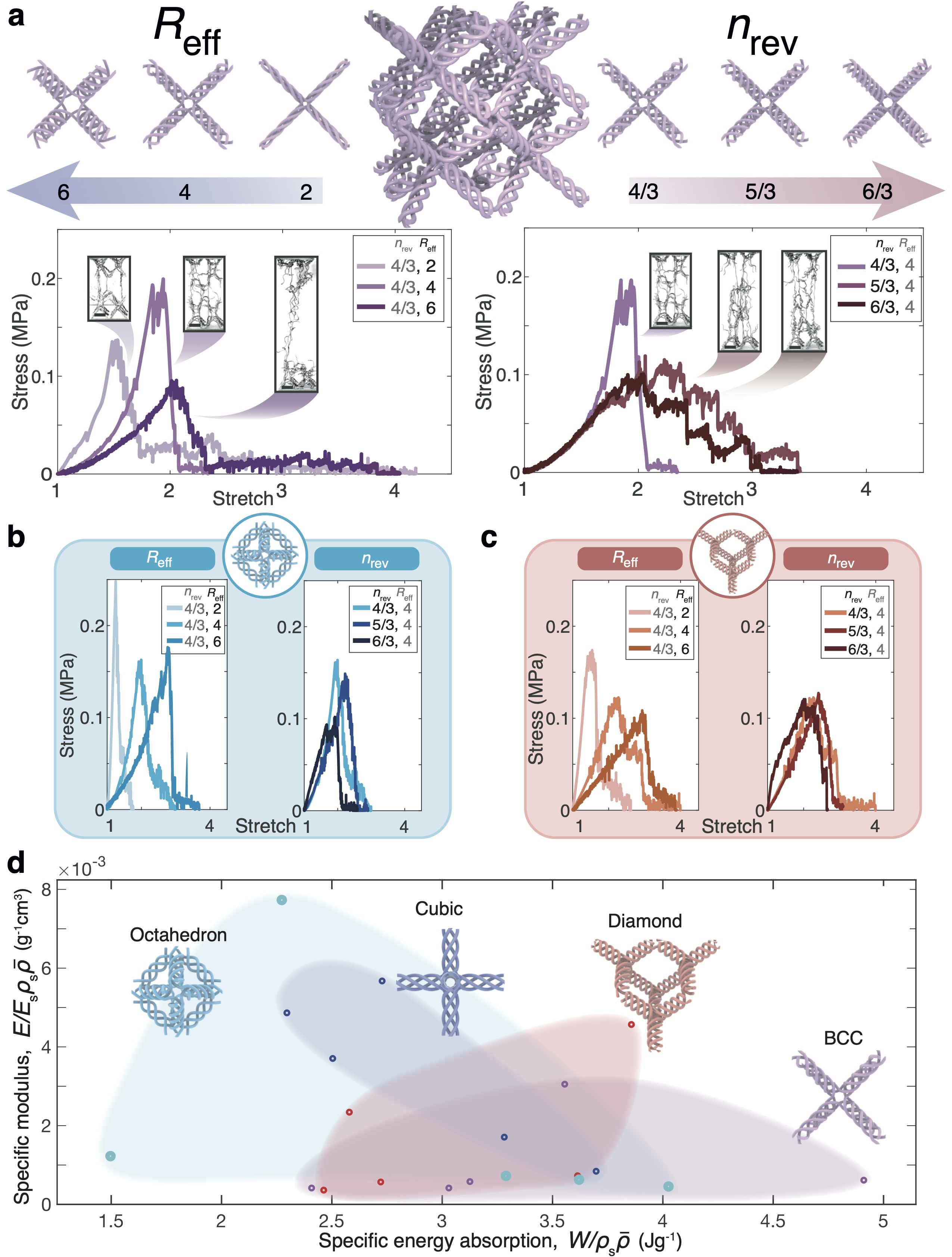}
\caption{\textbf{ \textit{In situ} tension experiments on woven lattices.} \textbf{a}, Experimental stress-stretch curves for 2$\times$2$\times$2 microscale BCC woven lattices under tension. Unit cell size, 60 \textmu{}m. Scale bar for inset SEM images, 40 \textmu{}m. \textit{Left:} Increasing $R_\mathrm{eff}$  reduces stiffness while increasing stretch at onset of failure.  \textit{Right:} Increasing $n_\mathrm{rev}$ significantly changes the peak stress and the characteristic shape of the stress-strain response.  \textbf{b}, Octahedron lattice. \textbf{c}, Diamond lattice. \textbf{d}, Specific modulus, i.e., effective modulus normalized by lattice density (defined as the product of the relative density $\bar{\rho}$ and the constituent material density $\rho_s$) and Young's modulus $E_s$ of the constituent polymer, versus specific energy absorption (normalized by lattice density) experimentally validates increased property space spanned by lattice topologies and corresponding geometric variations.}
\label{fig:3}
\end{figure}

\noindent\emph{Nonlinear Large-Deformation Response}
\hfill\\
\indent The primary utility of woven lattices lies far beyond their linear-elastic responses, in a large-deformation regime where they demonstrate exceptional deformability and resilience. To experimentally characterize the range of nonlinear responses available in woven lattices, we conducted \emph{in situ} uniaxial tension experiments on 2$\times$2$\times$2 tessellations of microscale woven lattices designed through our framework (Fig.~\ref{fig:3}). 
Using a two-photon lithography process with IP-Dip2 photoresist, each sample had unit cell sizes of $L = 60$ \textmu{}m and fiber radii of 1 \textmu{}m. The experiments were performed \emph{in situ} within a scanning electron microscope using a displacement-controlled nanoindenter (Alemnis AG) equipped with a custom tension gripper, ensuring precise control and real-time observation of stress-strain responses, deformation, and failure mechanisms (Fig.~S6). 
We designed these experiments to directly assess the role of design parameters such as effective radius $R_{\text{eff}}$ and fiber revolutions $n_{\text{rev}}$ on the large-deformation responses of these architectures.

In accordance with the linear perturbation simulations (Fig.~\ref{fig:linear}b), the experiments confirmed a reduction in the lattice stiffness as $R_{\text{eff}}$ increased (Fig.~\ref{fig:3}a to c, Fig.~S7). For example, reducing the $R_{\text{eff}}$ from 6 to 2 \textmu{}m for a BCC woven lattice of unit cell size $L = 60$ \textmu{}m results in an increase in linear stiffness from 40 kPa to 160 kPa. However, this comes at the cost of reduced stretchability before onset of failure (with failure stretches decreasing from $\sim$2 to 1.5), as the shorter fiber segments limit the available deformation before full uncoiling. The $R_{\text{eff}}$ parameter also determined the topology-dependent maximum stress prior to failure---e.g., providing the highest strength in the woven BCC lattice at $R_{\text{eff}}=4$ \textmu{}m, in contrast to $R_{\text{eff}}=2$ \textmu{}m for octahedral and diamond woven lattices.

The number of fiber revolutions per strut $n_{\text{rev}}$ also significantly alters the large-deformation responses, again leading to topology-dependent effects. In the woven BCC lattice, increasing $n_{\text{rev}}$ enhances ductility, leading to higher ultimate stretch and more gradual failure response. For instance, at $n_{\text{rev}} = 4/3$, failure occurs abruptly with a sharp drop in stress. However, at $n_{\text{rev}} = 5/3$ or $6/3$, stress decreases more gradually, indicating a transition from brittle-like failure to a more ductile response. This behavior can be attributed to increased fiber length, which allows for greater energy dissipation before rupture. In contrast, for octahedral and diamond-based woven lattices, the influence of $n_{\text{rev}}$ is less pronounced, likely due to fiber-topology-dependent entanglements, which can introduce localized stress concentrations that counteract the expected increase in ductility.

Overall, the results from experiments confirm that woven lattices provide an expansive mechanical property space, validating our design framework’s capability in creating lattices with a wide range of mechanical properties. While linear stiffness is primarily dictated by $R_{\text{eff}}$, large-strain behavior is strongly influenced by both $R_{\text{eff}}$ and $n_{\text{rev}}$, demonstrating the tunability of mechanical response through geometric variation. Notably, unlike monolithic beam lattices that typically fail at low stretch values ($\lambda<1.5$)\cite{zheng2016multiscale, montemayor2016insensitivity, mateos2019discrete, bauer2015push, hu2024compressive}, all tested woven lattices retained remarkable stretchability, sustaining stretches of at least 2 and up to 4 before complete failure. Additionally, failure strain consistently increased with $R_{\text{eff}}$, likely due to the greater available fiber lengths enabling more progressive deformation.

Furthermore, we examine the relation between specific modulus and specific energy absorption to assess the mechanical performance of woven lattices (Fig.~\ref{fig:3}d). Specific energy absorption is defined as the area under the stress-stretch curve, $W$, normalized by the lattice density. Through this analysis, we experimentally validated the significantly expanded material property space enabled by woven lattices. Importantly, even with just four fundamental lattice topologies, we observe a vast range of mechanical responses, underscoring the potential for further design exploration. This range suggests that by leveraging additional topological variations, such as the inclusion of more architectures, variation in sub-unit-cell parameters, and creation of spatially graded lattices, the mechanical properties of woven lattices can be tailored for a diverse array of applications, from energy-absorbing structures to highly flexible metamaterials.

\noindent\emph{Nonlinear Modeling Framework}
\hfill\\
\indent To gain mechanistic insight on the nonlinear, large-deformation responses of the woven lattices---as well as predictive capabilities of their performance---accurate computational modeling tools are required. While finite element analysis with continuum elements provides the highest-fidelity discretization (Fig.~S7), its high computational cost limits feasibility of large-tessellation modeling. Conversely, homogenization at the unit cell level fails to capture the intricate behavior of pseudoperiodic, functionally graded lattices, falling short in predicting the responses of parametric designs enabled by our design framework. 

\begin{figure}[H] 
\centering
\includegraphics[width=.9\textwidth]{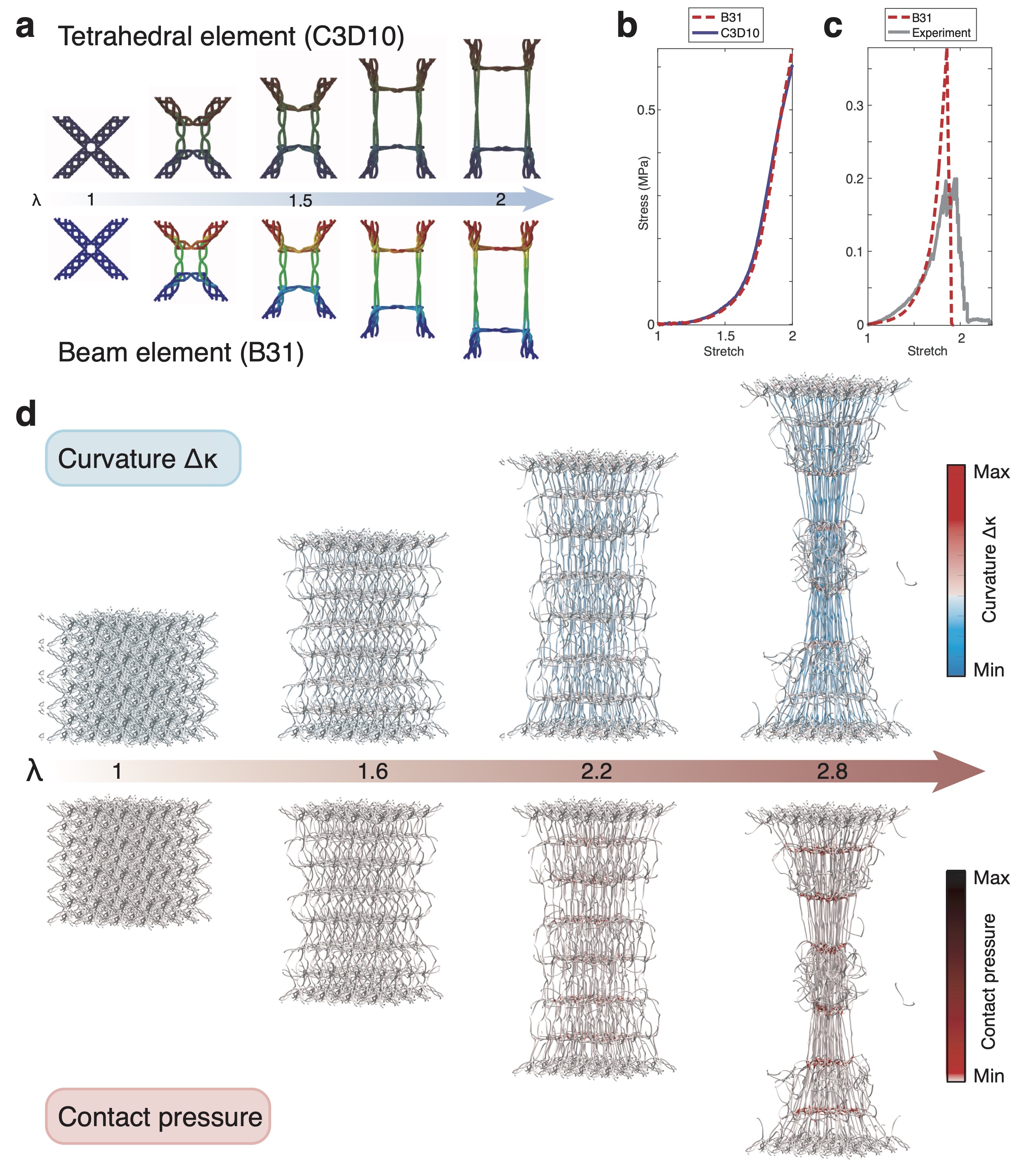}
\caption{\textbf{Finite-element modeling of large-deformation behavior.}  \textbf{a}, Comparison between quadratic tetrahedral- and beam-element representations of a BCC unit cell, with beam-element models demonstrating kinematic agreement and a four-order-of-magnitude reduction in computational cost. 
\textbf{b},  Stress-stretch responses of tetrahedral-element and beam-element simulations of a BCC unit cell, demonstrating agreement to a stretch of 2. \textbf{c}, Comparison between beam-element simulation and an experiment for a 2$\times$2$\times$2 tessellation of a BCC lattice. 
\textbf{d}, 
Simulation-enabled extraction of curvature and entanglement metrics on 4$\times$4$\times$4 woven-lattice tessellations. (Top) Change in curvature $\Delta\kappa$ starting from the undeformed configuration demonstrates fiber straightening and entanglement formation at nodal regions. (Bottom) The increasing effects of entanglement at high stretch are visible through the concentration of contact pressure at tightened nodes.}
\label{fig:4}
\end{figure}

To enable efficient large-tessellation simulations of these woven architectures, we employ Timoshenko beam-element representations that incorporate a Coulomb friction contact model as well as a strain-hardening material model (see Methods). This method is well-suited for simulating spatially complex, slender structures subjected to large strains. A comparison between beam-element and tetrahedral-element simulations reveals strong agreement in kinematic behavior and stress-stretch responses (Fig.~\ref{fig:4}a,b, Fig.~S9), indicating that the beam-element framework effectively captures the physics associated with their deformation. Comparing the experimental and simulated (beam-element representation) stress-stretch response of a BCC woven lattice also shows that the simulations can predict both the linear and nonlinear responses obtained from the experiments (Fig.~\ref{fig:4}c, Fig.~S8). Notably, the stretch at the onset of failure is accurately predicted by the simulations, while their higher peak stresses and more abrupt failure can be explained by the absence of manufacturing defects in our models which are inevitable in experimental specimens. In the experiments, these defects result in a distribution of failure strains within the woven fibers, leading to a more sequential or gradual failure, rather than the more abrupt failure observed in the simulations. The beam-element approach drastically reduces computational cost compared to the tetrahedral models (from 125.7 hours to 1 minute and 30 seconds on the same hardware, for the unit cell in Fig.~\ref{fig:4}a), making it feasible to simulate large tessellations. 

To demonstrate the applicability of the computational framework for larger woven-lattice tessellations, a 4$\times$4$\times$4 BCC woven-tessellation was simulated to a stretch of $\lambda = 4$ using the beam-element approach (Fig.~\ref{fig:4}d), revealing key mechanisms governing large-deformation behavior. As the lattice deforms, fibers initially straighten, as indicated by a negative change in curvature $\Delta \kappa$ (defined as the norm of the second derivative along the centerline) with respect to the initial configuration. At the same time, the models identify curvature increases which become highly concentrated in the entangled regions. Likewise, contact pressure, generated by entanglements between the woven fibers, localizes at these regions. 
The change in curvature characterizes the bending behavior of the fibers, directly influencing global deformation patterns, while contact pressure highlights regions of concentrated frictional interactions that govern load transfer and energy dissipation (Fig.~S10). Together, these parameters identify stress distribution pathways, revealing potential failure sites and high-energy dissipation zones. Understanding these mechanisms can enable the design of optimized fiber entanglements that enhance stretchability and toughness. Leveraging our computational framework to analyze the large-deformation behavior and mechanics of woven lattices, these findings highlight fiber topology as a crucial design parameter for tailoring and optimizing mechanical performance.

\begin{figure}[H] 
\centering
\includegraphics[width=\textwidth]{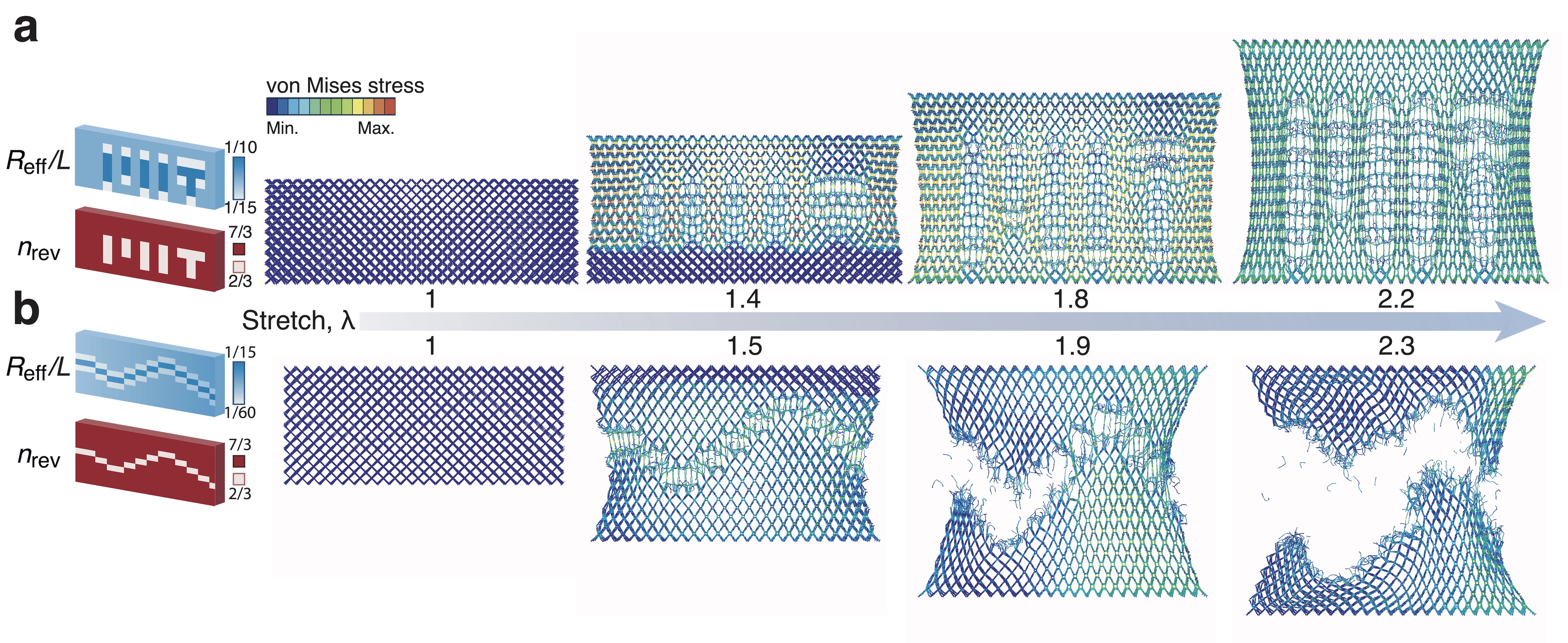}
\caption{\textbf{Engineered responses enabled by parametric design.} \textbf{a}, Targeted variation in $R_{\mathrm{eff}}$ and $n_\mathrm{rev}$ creates patterned deformation in woven architecture---depicting the letter pattern ``MIT''. \textbf{b}, This variation can also be used to design defects in order to architect failure.}
\label{fig:5}
\end{figure}

\noindent\emph{Tailored Deformation and Failure Patterns}
\hfill\\
\indent The proposed design framework enables the creation of woven lattices with precisely tailored deformation and failure behaviors (Fig.~\ref{fig:5}). As an example, Fig.~\ref{fig:5}a demonstrates a 30$\times$10$\times$2 BCC woven-lattice-sheet engineered to form a visually distinct text pattern under large strains. This effect is achieved by assigning different unit-cell parameters to the background, boundary, and text-pattern regions, creating a mechanically tunable system that enhances visual contrast through controlled deformation. The woven lattice in the text region is designed with lower $n_{\mathrm{rev}}=2/3$ and higher $R_{\mathrm{eff}}/L=1/10$ (where $L$ corresponds to the unit cell size), resulting in a highly compliant architecture with comparatively low entanglement that reshapes visibly under strain. This early deformation causes the fibers to straighten and the nodes to tighten, increasing local stretch and making the letters visually prominent. The background region, i.e., the lattice unit cells outside of the text region, is instead assigned $R_{\mathrm{eff}}/L=1/12$ and $n_{\mathrm{rev}}=7/3$, creating a stiffer network that resists deformation, ensuring that the surrounding material provides a contrasting frame for the stretching text region. To further refine the deformation contrast and maintain pattern clarity, a boundary zone is introduced at the interface between the text and background regions. This boundary region is assigned $R_{\mathrm{eff}}=1/15$ and $n_{\mathrm{rev}}=7/3$, resulting in the lowest compliance. By deforming the least, these unit cells compensate for the large elongation of the text region unit cells and reduce blurring of the pattern at the edges of the text region.

Beyond controlled deformation, this framework also enables the strategic design of failure pathways. As shown in Fig.~\ref{fig:5}b, a 24$\times$10$\times$2 woven lattice is patterned with a sinusoidal failure path using the similar design principles as those in the letter pattern. Additionally, a linear gradient in effective radius is introduced, reducing $R_{\mathrm{eff}}$ towards the chosen end of the lattice. This variation concentrates stress and directs failure propagation in a controlled manner, with failure initiating at the end with smaller radii due to the lower available length of fibers on that side. The gradient follows a stepwise linear reduction in $R_{\mathrm{eff}}$ at each unit cell boundary (up to 1/20 over the width of the lattice).
This controlled spatial variation in fiber topology enables predictable failure pathways, demonstrating how the woven lattice framework can be leveraged to program both deformation and fracture behavior.

\subsection{Discussion}
\hfill\\
The presented generalized design framework enables the seamless creation of arbitrarily complex 3D woven architected materials with highly tunable mechanical responses. By enabling precise control over geometric parameters across multiple length scales---both within individual unit cells and across a tessellated structure---this framework allows for the rapid design of spatially graded and topologically heterogeneous materials. Unlike conventional CAD-based approaches that require manual adjustments, this method significantly streamlines the design of complex woven architectures. Through nonlinear finite element simulations and microscale \emph{in situ} tension experiments, we have demonstrated how key parameters dictate variations in mechanical behavior, enabling tunable stiffness, extreme anisotropy (exceeding $E_{\max}/E_{\min}$ = 20), ultralarge stretchability (up to a stretch of 4), and enhanced energy dissipation through frictional contact and fiber entanglement. These results underscore the vast design space available for engineering compliant woven networks with programmable mechanical properties. While this study focuses on five fundamental woven unit-cell topologies, future extensions of this framework could explore multi-scale hierarchical lattices that integrate stochastic and periodic elements to achieve tailored mechanical responses. The incorporation of active or stimuli-responsive fibers could further enable adaptive woven lattices with dynamically reconfigurable properties. 

The potential applications of these tunable woven architectures span multiple disciplines. In biomedical engineering, woven scaffolds with tailored mechanical compliance could guide cell migration and tissue regeneration, while insights into fibrous network mechanics could help bridge the gap between synthetic and biological materials, such as cytoskeletal networks~\cite{wang_long-range_2014}. In materials science, synthetic woven architectures provide opportunities for defect-tolerant materials, where distributed load-sharing and entanglement-driven stress redistribution enhance mechanical resilience~\cite{wang_structured_2021}. In soft robotics, architected entanglement offers a pathway to stretchable, damage-tolerant actuators, while in aerospace and defense, energy-absorbing woven materials could serve as lightweight, impact-mitigating layers. By advancing the design and fabrication of woven lattices, this framework establishes a foundation for next-generation architected materials with highly programmable mechanical properties. Continued refinement of these design strategies will further expand the utility of woven networks, enabling new possibilities for functional and high-performance materials across scientific and engineering disciplines.

\subsection{Methods}
\hfill\\
\emph{Selection of Geometric Parameters}
\hfill\\
Geometric parameters in experiments were varied within bounds determined by manufacturability considerations. All samples were fabricated with a unit cell size $L$ of 60~\textmu{}m (except the more spatially complex tetrakaidecahedron, which was printed with a unit cell size of 120~\textmu{}m), to obtain a well-defined unit cell while minimizing lattice stitching required by workspace limits. The effective radius $R_{\mathrm{eff}}$ was varied between 2 and 6 ~\textmu{}m, leading to $R_{\mathrm{eff}}/L$ ratios between $1/30$ and $1/10$. The number of turns  $n_{\mathrm{rev}}$, limited by manufacturable strand spacing, ranged between $\frac{4}{3}$ and $\frac{7}{3}$ (for lattices consisting of 3-helices: diamond, octahedron, BCC, and tetrakaidecahedron) and $\frac{4}{4}$ and $\frac{6}{4}$ for the 4-helix cubic lattice. Fiber radius was fixed at 1~\textmu{}m. Relative densities $\bar{\rho}$ ranged between 1\% and 5\%, determined by the spatial density of the generating lattice and fiber packing density as determined by both $R_{\mathrm{eff}}$ and $n_{\mathrm{rev}}$. To compare these manufactured lattices, the effective stiffness and the energy absorption capacity (computed as the area under the stress-strain curve to a set strain) were normalized by relative density $\bar{\rho}$ and the density of the constituent IP-Dip2 polymer $\rho_s$. 

\noindent\emph{Fabrication and Mechanical Characterization of Microscale Lattices} 
\hfill\\
Representative 2$\times$2$\times$2 lattice samples were fabricated on a silicon substrate using two-photon lithography (Photonic Professional GT2, Nanoscribe GmbH) out of IP-Dip2, an acrylate-based photoresist. The architected structures were printed at 25 mW laser power with a 10 mm s$^{-1}$ scan speed, while monolithic tensile fixtures were fabricated at 20 mW to reduce cavitation. Printing parameters remained consistent, with a uniform hatching and slicing distance of 0.2 \textmu{}m for lattices and pillars and 0.3 \textmu{}m for the tensile fixtures. Post-processing involved immersion in propylene glycol monomethyl ether acetate for up to 2 hours to remove residual resin, followed by a 5-minute isopropanol rinse and drying via critical point drying (Autosamdri 931, Tousimis). To enhance visualization for \emph{in situ} mechanical characterization, a 10 nm gold coating was applied using sputtering (SCD 040, Balzers).

For mechanical characterization, uniaxial tension experiments were performed using a custom gripper mounted on a displacement-controlled nanoindenter (Alemnis AG) inside a scanning electron microscope (Gemini 450, ZEISS), enabling real-time imaging. Specimens were loaded at a strain rate of 5$\times$10$^{-2}$ s$^{-1}$. Stress-strain data were derived from load-displacement curves, normalized by nominal cross-sectional area and height. The dissipated energy density was determined by integrating the area under the stress-stretch curve.

\noindent\emph{Finite Element Modeling}
\hfill\\
To model complex large-deformation responses, finite element models employing beam elements were conducted using ABAQUS/Explicit 2023. The centerline geometry was directly used to create meshes using linear Timoshenko beam elements (B31) with general contact using an isotropic Coulomb friction model with an empirically matched dynamic friction coefficient of 0.5 and a hard-contact criterion in the normal direction. Failure was modeled through element deletion based on a maximum strain criterion of .0377. The material model consisted of an isotropic elastic-plastic model experimentally derived from the experimental response of an IP-Dip2 micropillar under uniaxial tension. 

Linear perturbation simulations were performed using nTop (nTop Inc.) and in ABAQUS/Standard; an orthogonal set of linear perturbations was performed with periodic boundary conditions to obtain the full elasticity and compliance tensors for each unit cell. For these homogenization simulations, periodic tetrahedral meshes were used, generated using the open source finite element mesh generator Gmsh~\cite{geuzaine2009gmsh}. Mesh convergence (see Supplementary Information) suggested a consistent mesh size of .25 of the fiber radius, generating meshes with approximately $10^5$--$10^6$ quadratic tetrahedral elements (C3D10). Timing comparison (Fig.~\ref{fig:4}a) was performed on Intel\textsuperscript{\textregistered} Xeon\textsuperscript{\textregistered} W-2245 CPU @ 3.90GHz with 8 cores and 128GB of RAM.

\subsection{Data availability}
\hfill\\
The data that support the findings of this study are present in the paper and/or in the Supplementary Information. Additional data related to the paper are available from the corresponding author upon request.

\subsection{Acknowledgements}
\hfill\\
C.M.P. acknowledges support from the National Science Foundation (NSF) CAREER Award (CMMI-214246) and CMMI-2418432. M.C. acknowledges support from the MIT School of Engineering Postdoctoral Fellowship Program for Engineering Excellence. This work was performed in part through use of MIT.nano's Fabrication and Characterization Facilities. The authors acknowledge the support of Jiayi Wu in packaging the design framework. 

\subsection{Author contributions}
\hfill\\
C.M.P. and M.C. conceived this study. M.C. designed, developed, and implemented the software. J.U.S. fabricated samples, performed mechanical experiments, and analyzed data. B.F.G.A., J.U.S., and M.C. performed simulations and analyzed data. C.M.P. supervised the project. M.C., J.U.S., and C.M.P. wrote the manuscript with input from all authors.

\subsection{Competing interests}
\hfill\\
The authors declare no competing interests.

\nolinenumbers
\vspace{10pt}

\newpage
\subsection{References} 
\hfill \\

\newpage

\title{Supplementary Information \hfill\\
Design framework for programmable three-dimensional woven metamaterials}
\author{Molly Carton$^{1,2}$, James Utama Surjadi$^{1}$, Bastien F. G. Aymon$^{1}$, Carlos M.~Portela$^{1\ast}$}

\maketitle

\begin{affiliations}
\footnotesize
 \item Department of Mechanical Engineering, Massachusetts Institute of Technology, Cambridge, MA 02139, USA
 \item Department   of   Mechanical   Engineering,   University of Maryland, College Park, MD, USA
\end{affiliations}
\vspace{10pt}
\spacing{1.0}

\subsection{Supplementary Text}
\hfill\\
\noindent\emph{S1: Woven topology and connectivity of fibers}

Effective woven beams in woven metamaterials are composed of intertwined helical fibers. To ensure no fibers cross, all helices and the connections between them have the same chirality. By maintaining a direction of twist, constituent fibers are able to wrap the node surface without crossing. We define the woven lattice  as depicted in Fig.~1, and in more detail, in Fig.~\ref{fig:fulloutline}. The corresponding ``monolithic" beam lattice forms a spatial graph consisting of interconnected vertices and edges (Fig.~\ref{fig:fulloutline}a). For each vertex in this graph, we find the spherical dual (Fig.~\ref{fig:fulloutline}b), which consists of vertices that each represent a ``face" of the polygonal subgraph at each node in the original graph, and edges that connect neighboring vertices. Each vertex in this dual is realized as an $n$-helix, where $n$ is the connectivity of the respective node; each edge is realized as a curve in space connecting one strand of the $n$-helix to a strand on the neighboring node. Connecting these strands with a tangent path between the helix bases (Fig.~\ref{fig:fulloutline}c) creates a chiral, non-intersecting network. This path, which in projection follows a geodesic of the nodal sphere (Fig.~\ref{fig:fulloutline}d), connects the helices with the shortest possible spherical distance. The radial shape of this path (Fig.~\ref{fig:fulloutline}e) is constructed as a B\'{e}zier curve to smoothly interpolate the helix angles ($\alpha$, with respect to the surface of the sphere).  By connecting each of the $n$-helix beams with these paths, the woven node is realized (Fig.~\ref{fig:fulloutline}f).

\noindent\emph{S2: Strand matching of nonuniform unit cells}
\hfill\\

The helices comprising a given effective beam are defined, on the surface of the nodal sphere, by a `small circle' of (euclidean/extrinsic) radius $R_\text{eff}$ and a vector $\vec{v_1}$ defining the direction from the center of the nodal sphere -- identically the direction of the beam centerline. To avoid crossing fibers, these helices must not overlap on the node -- this is prevented by allowing sufficient intrinsic (spherical) distance between the intersection of the nodal sphere with $\vec{v_1}$ and $\vec{v_2}$ = $d_\text{intrinsic} >R_\text{1,intrinsic}+R_\text{2,intrinsic}$. This determines the minimum radius of the nodal sphere $r$ such that  $d_\text{intrinsic} >R_\text{i,intrinsic}+R_\text{j,intrinsic}$ for all neighboring pairs $i$, $j$. The example in figure Figure~\ref{fig:nodes} demonstrates the geometric node adjustment to accomodate this varying radius. The central node, connecting between beams of smaller ($R_\mathrm{eff} = 1/30$) and larger ($R_\mathrm{eff} = 2/15$) radius, matches the size of the rightmost node ($R_\mathrm{eff} = 2/15$) in order to accomodate the spacing of the largest effective radii.

\newpage
\subsection{Supplementary Figures}
\renewcommand{\thefigure}{S\arabic{figure}}
\setcounter{figure}{0}   
\hfill\\
\begin{figure}[H] 
\centering
\includegraphics[width=\textwidth]{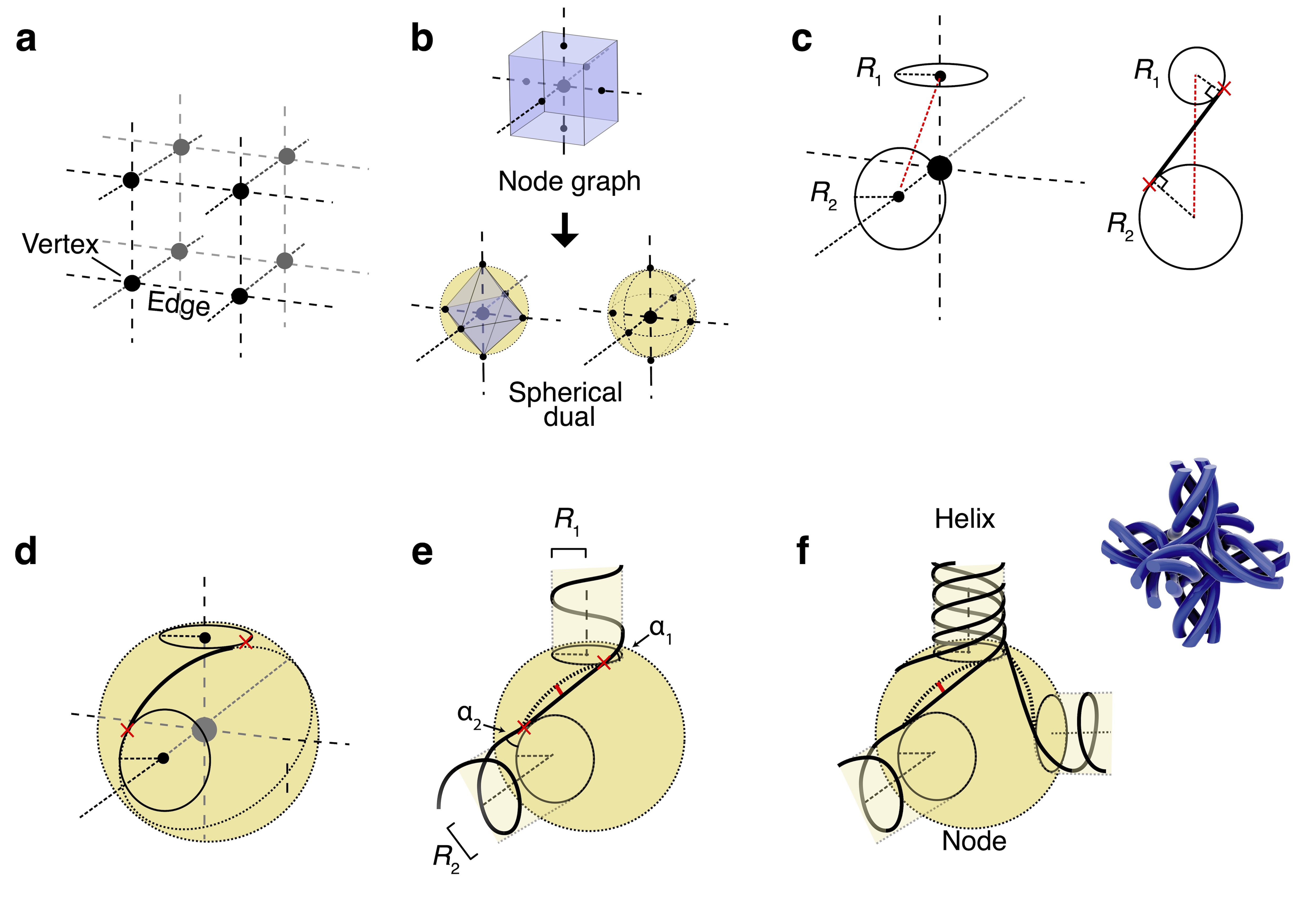}
\caption{\textbf{Construction process of the node in a woven lattice.} From \textbf{a}, the lattice topology graph (composed of edges and vertices), \textbf{b}, take the spherical dual of the lattice at a vertex to construct the node graph. Each of the vertices in this graph will be connected with a strand to form the fully connected woven node; connectivity here determines the number of strands on an effective beam. \textbf{c}, For each pair of nodes, construct the path tangent to the helix base circles of radius $R_{\text{eff},1}$ and $R_{\text{eff},2}$. \textbf{d}, This shortest path follows a great circle of the nodal sphere. The end points of this path (labeled with $\times$) are the endpoints of the helices forming the connected strands. \textbf{e}, Helix angles $\alpha_1$ and $\alpha_2$, determined by $n_\mathrm{rev}$, define the path of the B\'{e}zier curve that connects these helices.} 
\label{fig:fulloutline}
\end{figure}

\newpage
\begin{figure}[H] 
\centering
\includegraphics[width=.7\textwidth]{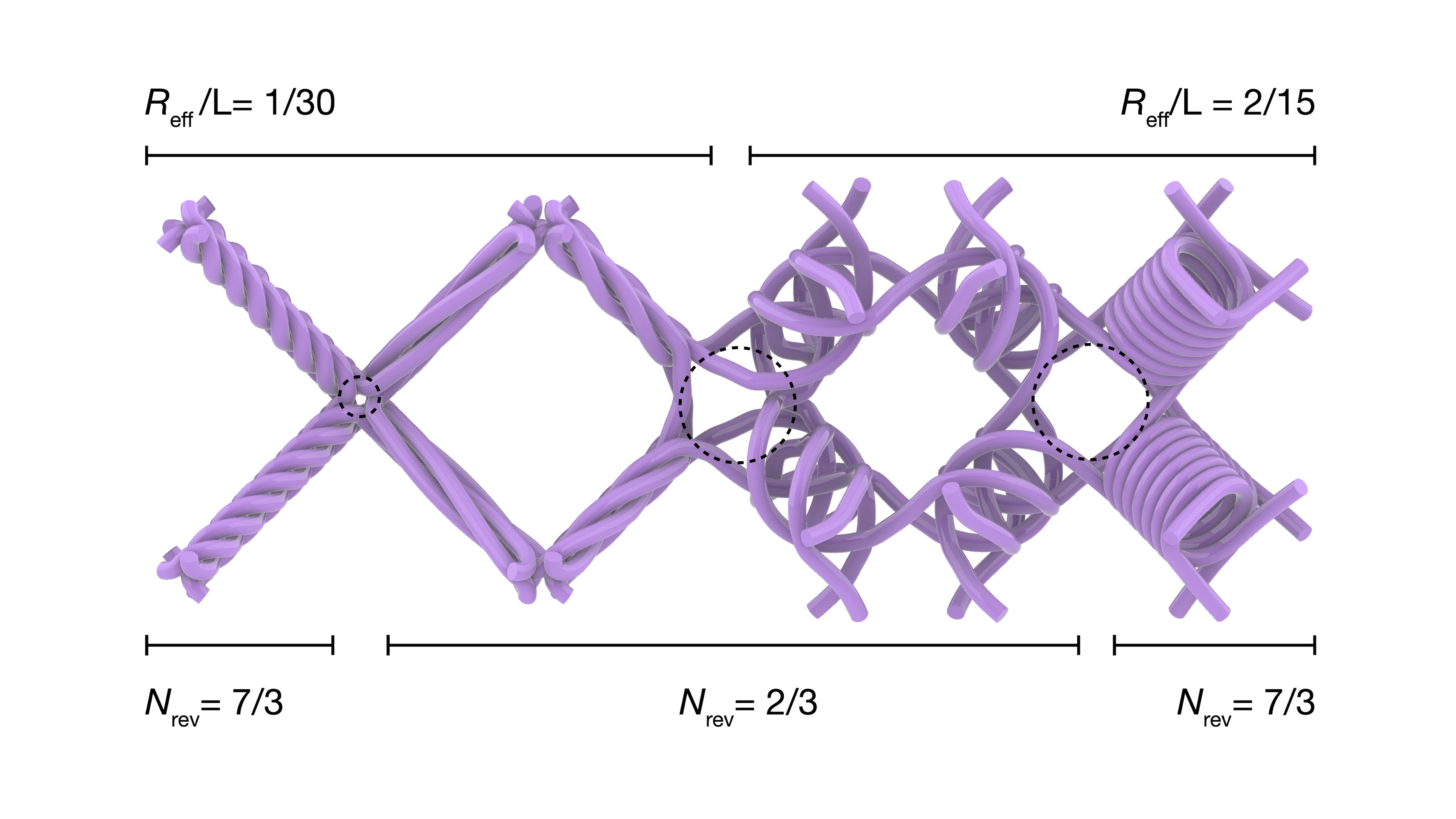}
\caption{\textbf{Construction of the node in lattice with nonuniform unit cells.} In order to maintain continuity and smooth curvature while at a node with varied $R_\mathrm{eff}$, the woven node radius (depicted by black dotted circles) must expand to accomodate the largest $R_\mathrm{eff}$ on the beams connecting at that node. Altering $n_\mathrm{rev}$ does not affect node radius; however, as shown in Fig~\ref{fig:fulloutline}e, the resulting pitch does affect the helix angle $\alpha$ and thus the connecting B\'{e}zier curve. }

\label{fig:nodes}
\end{figure}

\newpage
\begin{figure}[H] 
\centering
\includegraphics[width=\textwidth]{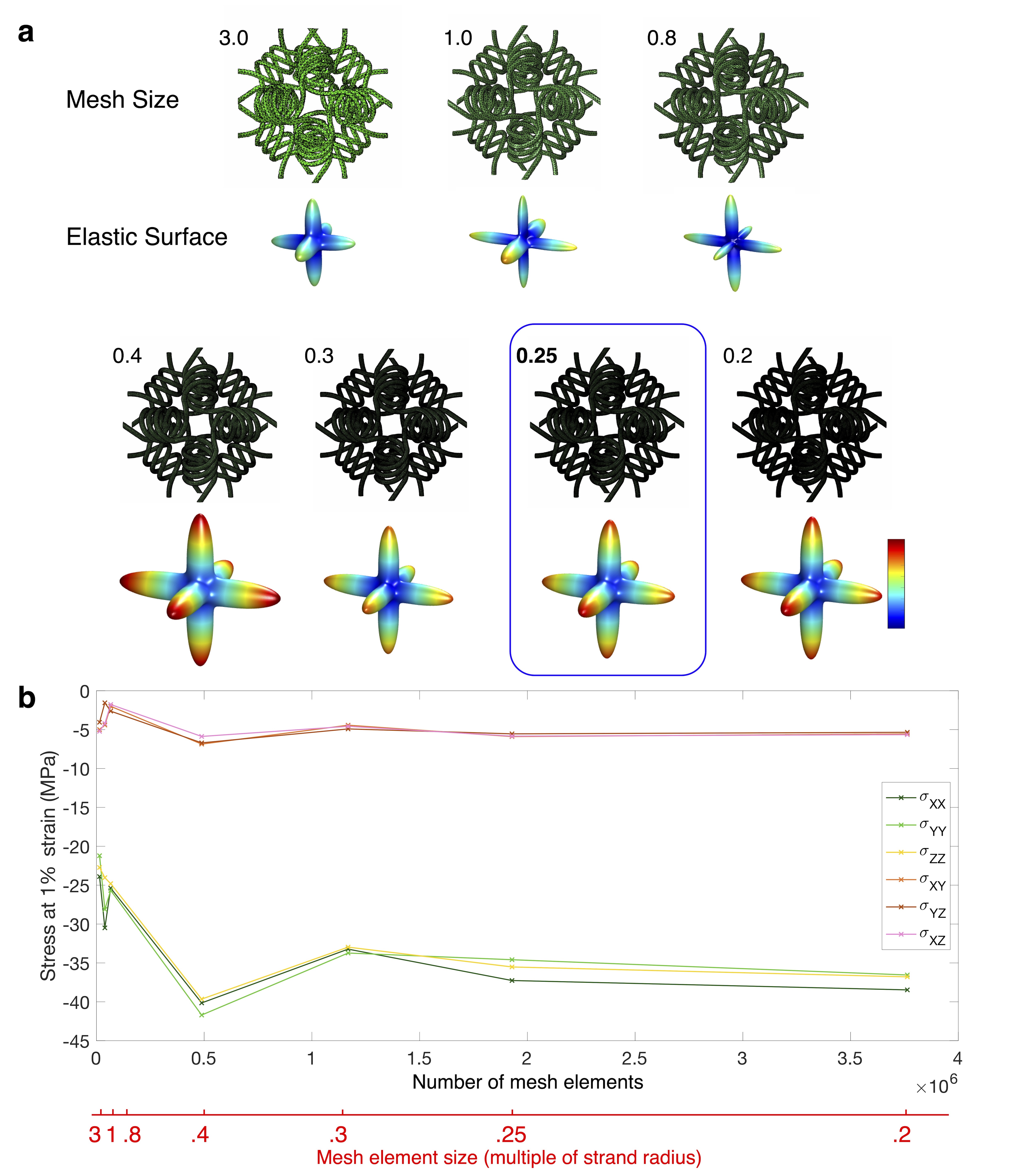}
\caption{\textbf{Mesh convergence study on effect of element size on linear stress values}. \textbf{a}, Mesh and resultant elastic surfaces. 
\textbf{b}, Stress due to applied 1\% strain with periodic boundary conditions. Cases $\sigma_{ij}$ represent applied strain via the  $\varepsilon_{ij}$ component. Periodic meshes generated using Gmsh~\cite{geuzaine2009gmsh} in tetrahedral elements (C3D10). }

\label{fig:periodic_fea}
\end{figure}

 \newpage
\begin{figure}[H] 
\centering
\includegraphics[width=\textwidth]{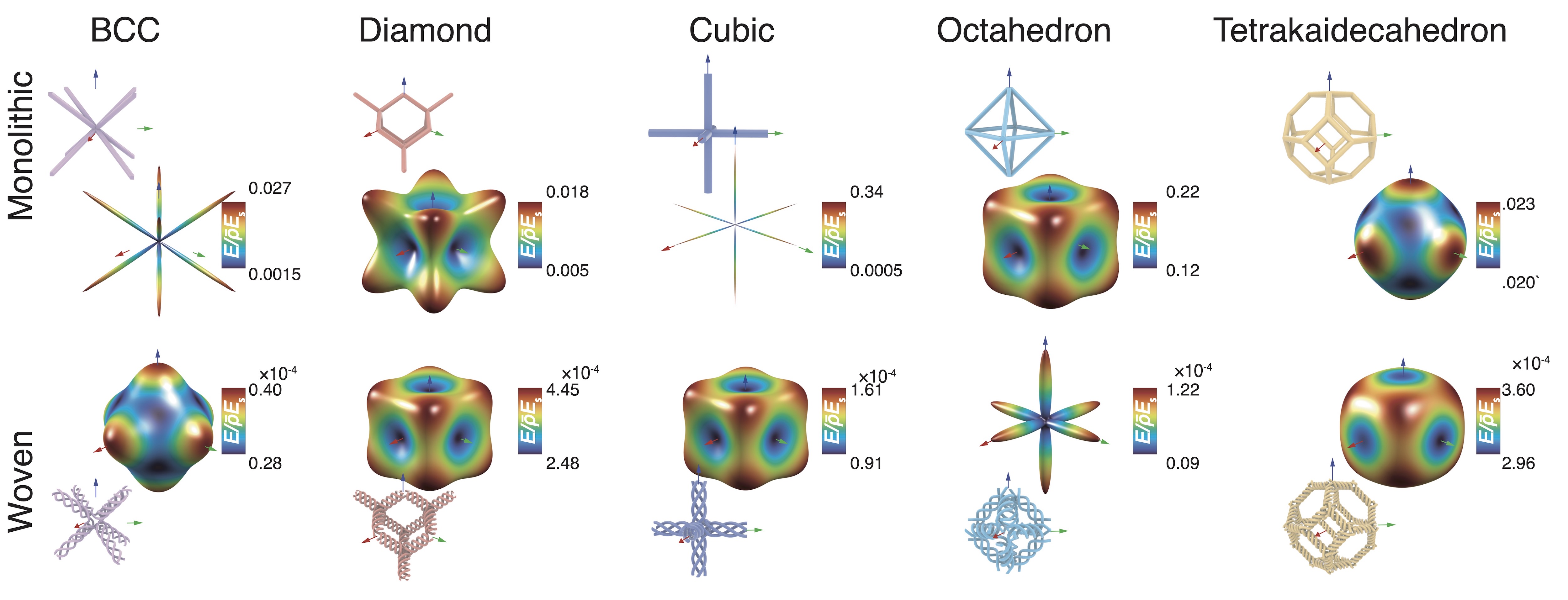}
\caption{\textbf{Comparison of elastic surfaces for monolithic beam lattices and woven lattices.} The elastic surfaces, from left, for BCC, diamond cubic, cubic, octahedron, and tetrakaidecahedron beam lattices compared to their woven counterparts. }
\label{fig:monolithic_woven}
\end{figure}

\newpage
\begin{figure}[H] 
\centering
\includegraphics[width=\textwidth]{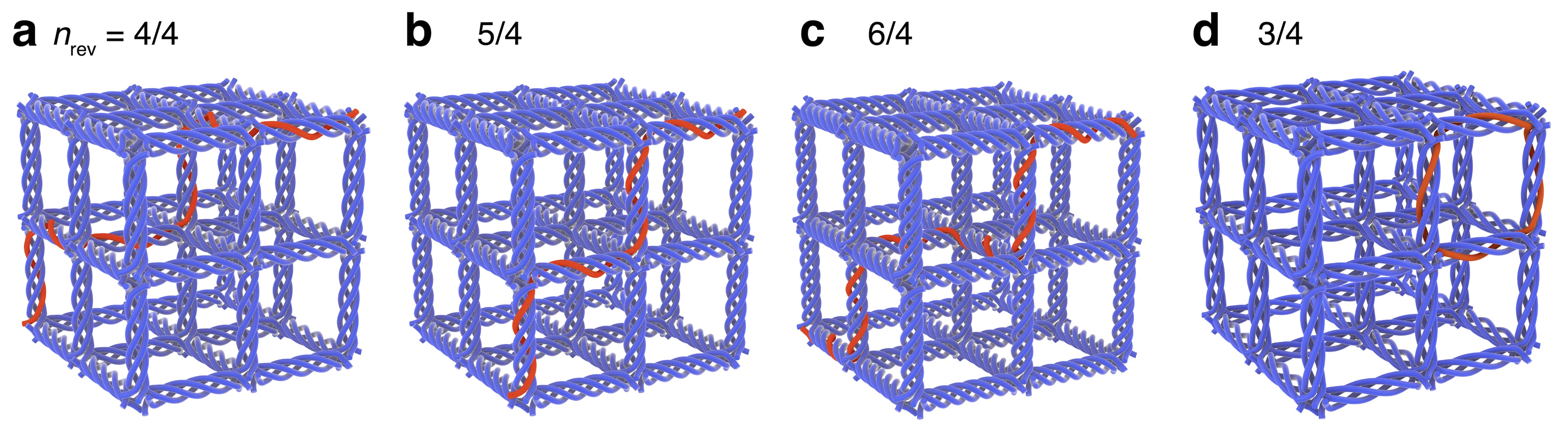}
\caption{\textbf{Varying $\mathbf{n_\mathrm{rev}}$ changes lattice-level topology.} The parameter $n_\mathrm{rev}$ affects the path of a strand along a lattice beam, modifying how strands connect unit cells within a lattice. The cubic lattice is of connectivity degree 4 (i.e., each beam is realized as a 4-helix). Each integer value of ($\frac{n_\mathrm{rev}}{4}) \mod 4$ (= 0, 1, 2, 3) produces a different strand path through the 2x2x2 lattice: \textbf{a--c}, Values of $n_\mathrm{rev} = 4/4, 5/4, 6/4$: ($\frac{n_\mathrm{rev}}{4}) \mod 4$ = 0,1,2 produce strand paths that traverse the lattice along different paths.  
\textbf{d}, A value of $n_\mathrm{rev} = 3/4$: ($\frac{n_\mathrm{rev}}{4}) \mod 4$ = 3 causes the strand to link into a ring. The resulting lattice consists of a chain-mail-like interlocking structure.}

\label{fig:fiberpathchange}
\end{figure}

\newpage
\begin{figure}[H] 
\centering
\includegraphics[width=.8\textwidth]{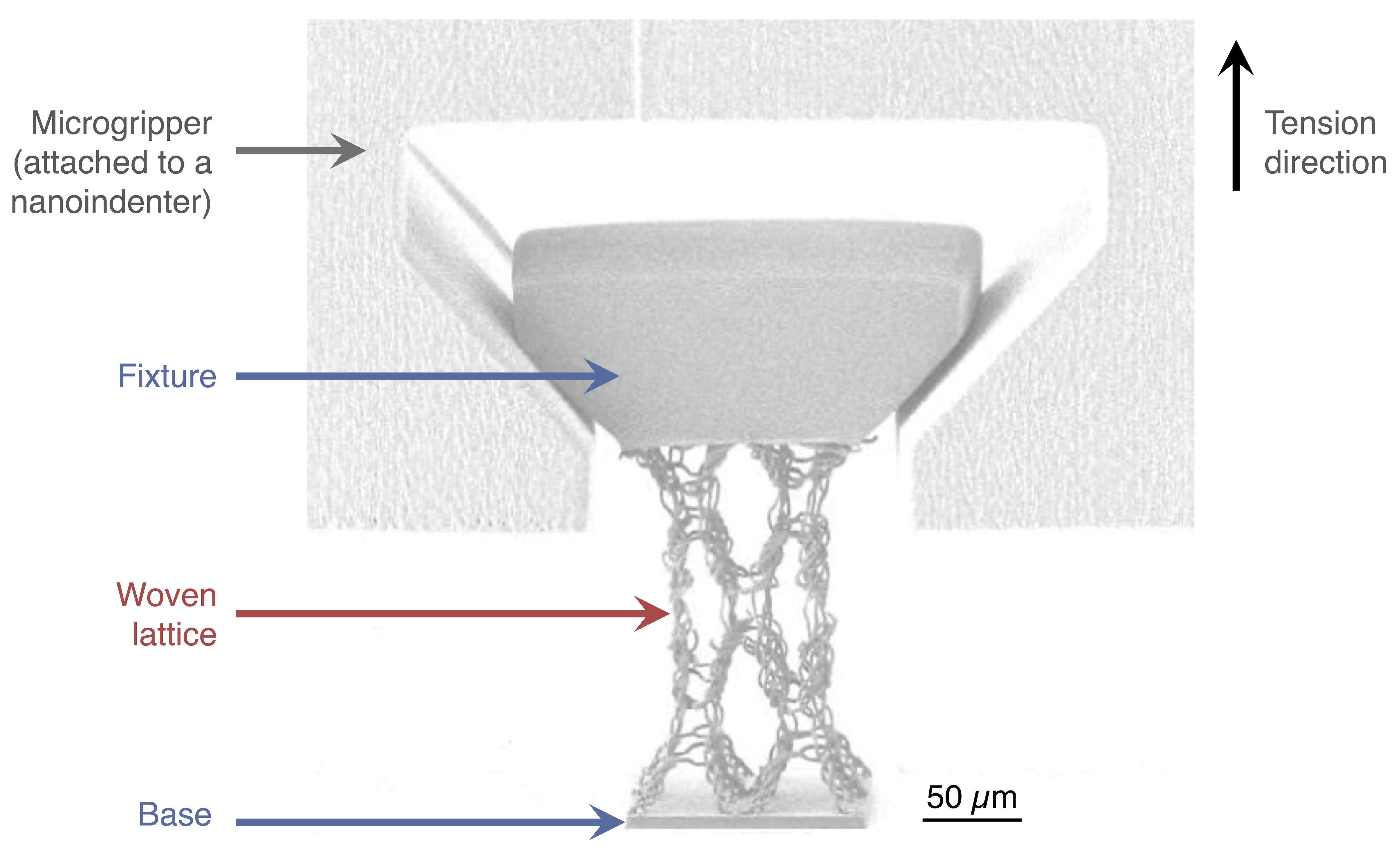}
\caption{\textbf{Experimental setup for \textit{in situ} tension of woven lattices.} From top, custom printed tensile microgripper mounted to nanoindenter, monolithic printed tensile fixture, woven lattice, base attached to silicon print substrate.}
\label{fig:experimentalsetup}
\end{figure}

\newpage
\begin{figure}[H] 
\centering
\includegraphics[width=\textwidth]{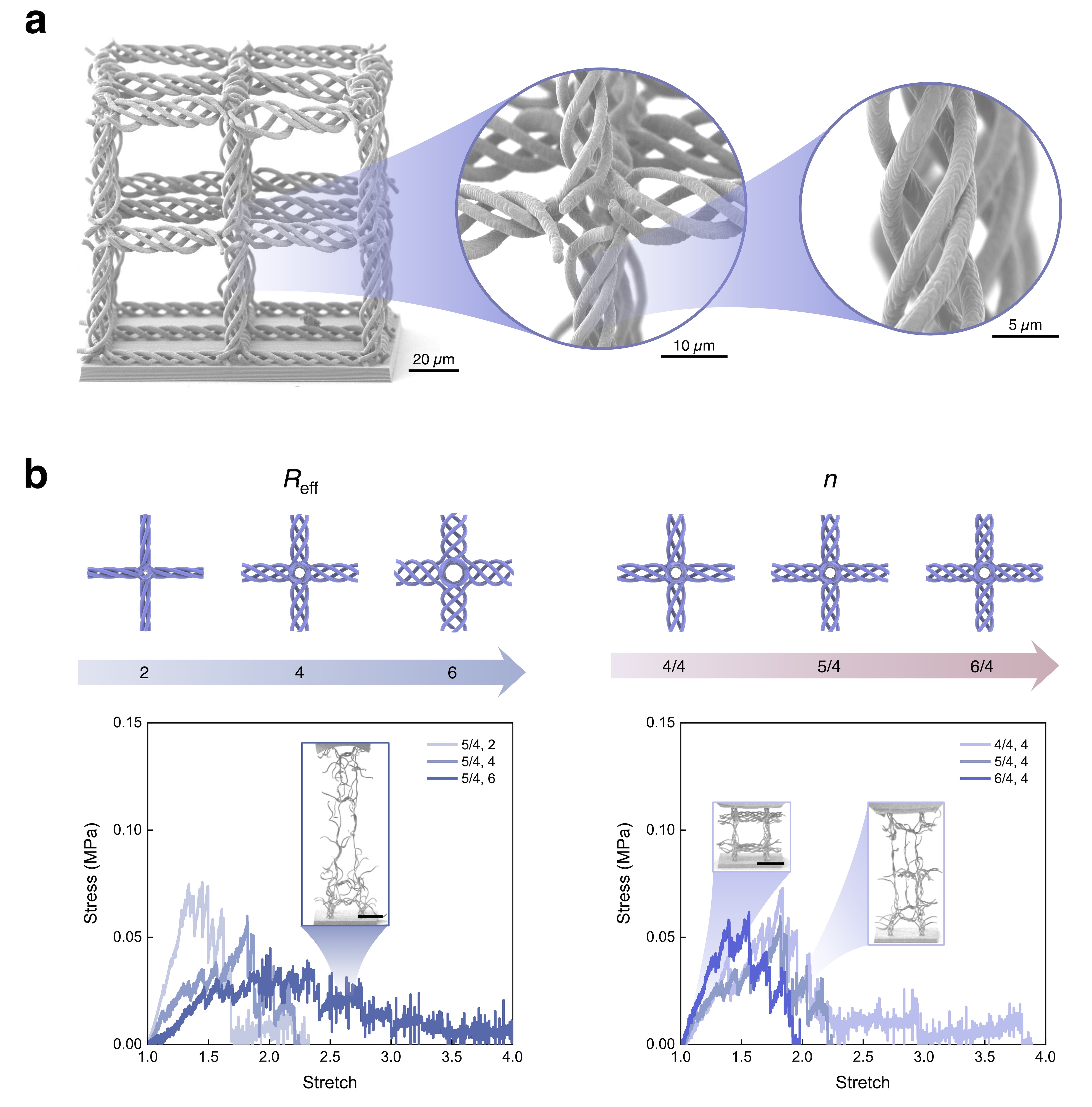}
\caption{\textbf{\textit{In situ} experimental tension of cubic woven lattices.} \textbf{a}, SEM images of a fabricated 2$\times$2$\times$2 cubic lattice.
\textbf{b}, Experimental stress-stretch curves for 2$\times$2$\times$2 micro-fabricated cubic woven lattices under tension. Scale bars in \textbf{b}, 50 \textmu{}m.}
\label{fig:cubicexperiments}
\end{figure}

\newpage
\begin{figure}[H] 
\centering
\includegraphics[width=\textwidth]{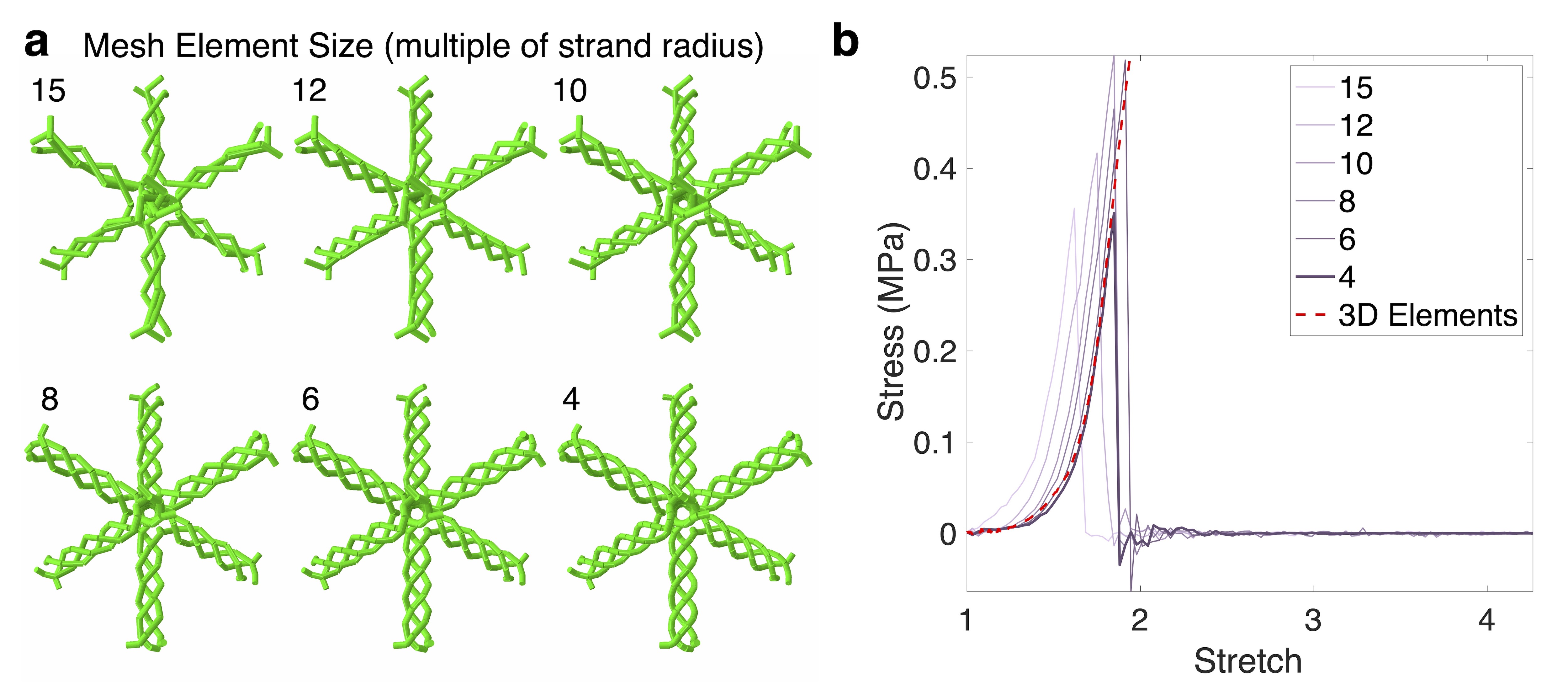}
\caption{\textbf{Mesh convergence study on effect of beam element size on stress-stretch results for a unit cell.} \textbf{a}, Mesh sizes. \textbf{b}, Resulting stress-stretch curves, with 3D tetrahedral-element results up to a stretch of 1.
}
\label{fig:tetsvsbeam}
\end{figure}

\newpage
\begin{figure}[H] 
\centering
\includegraphics[width=\textwidth]{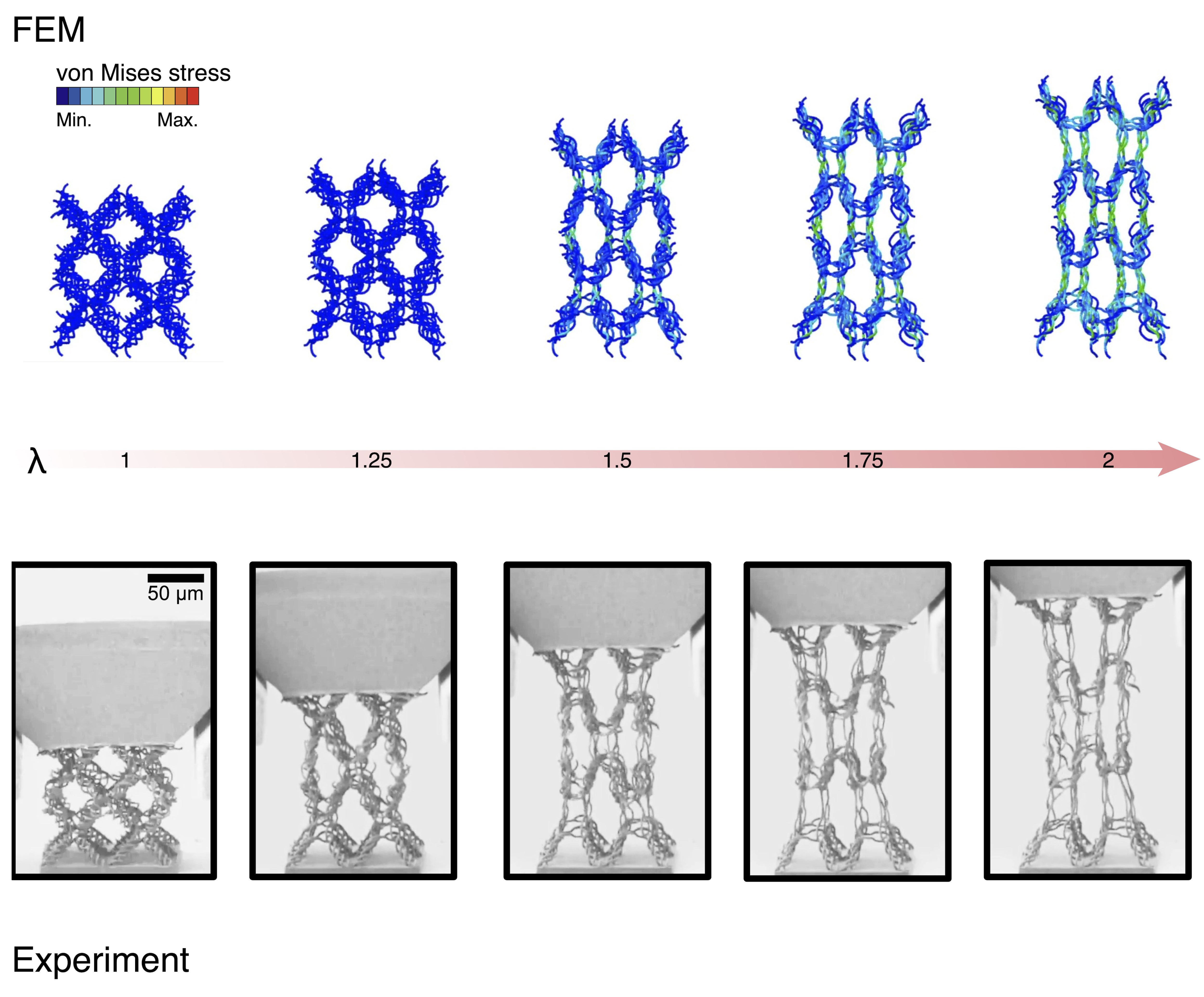}
\caption{\textbf{Side-by-side comparison of experimental results and beam-based finite element modeling (FEM) of woven lattices under tensile loading.} Beam-element behavior kinematics matches visually with in situ experiments (see Supplementary Video 1). Scale bar, 50 \textmu{}m. }
\label{fig:experimentsvsfem}
\end{figure}

 \newpage
\begin{figure}[H] 
\centering
\includegraphics[width=\textwidth]{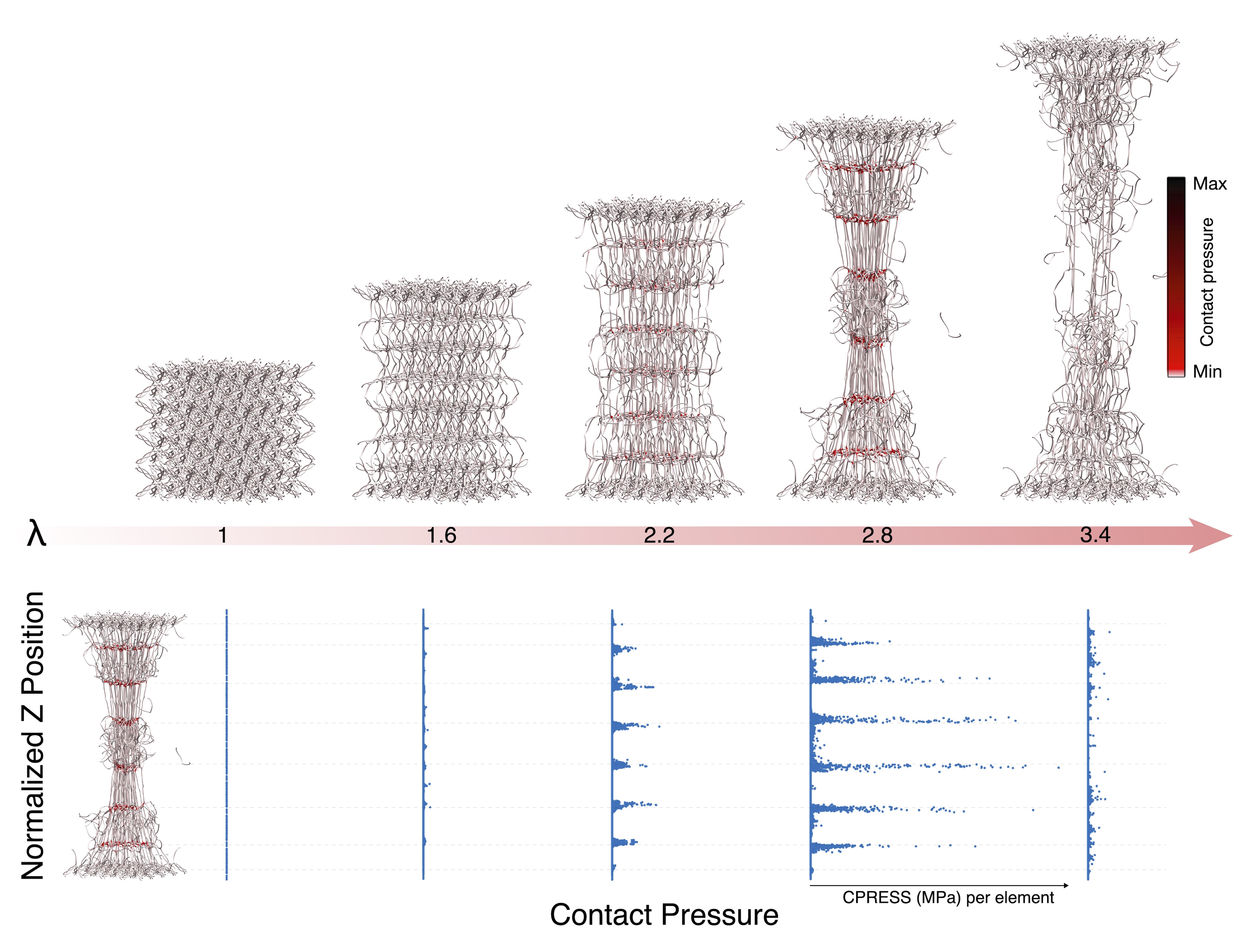}
\caption{\textbf{Evolution of contact pressure in beam-element simulation of 4$\times$4$\times$4 BCC lattice (Fig. 4).} Contact begins at a stretch of about 1.6, at which point contact pressure appears at the nodes between and in the center of each unit cell. This pressure increases as entanglement causes knotting of fibers at these nodes and straightening of the intervening fibers; the contact pressure is relatively consistent at each of these nodes, reaching maximum far from the boundaries where deformation is greatest. Failure begins at a stretch of about 3, after which contact pressure drops and delocalizes as failed strands slip. }
\label{fig:cpress}
\end{figure}

\newpage
\begin{figure}[H] 
\centering
\includegraphics[width=\textwidth]{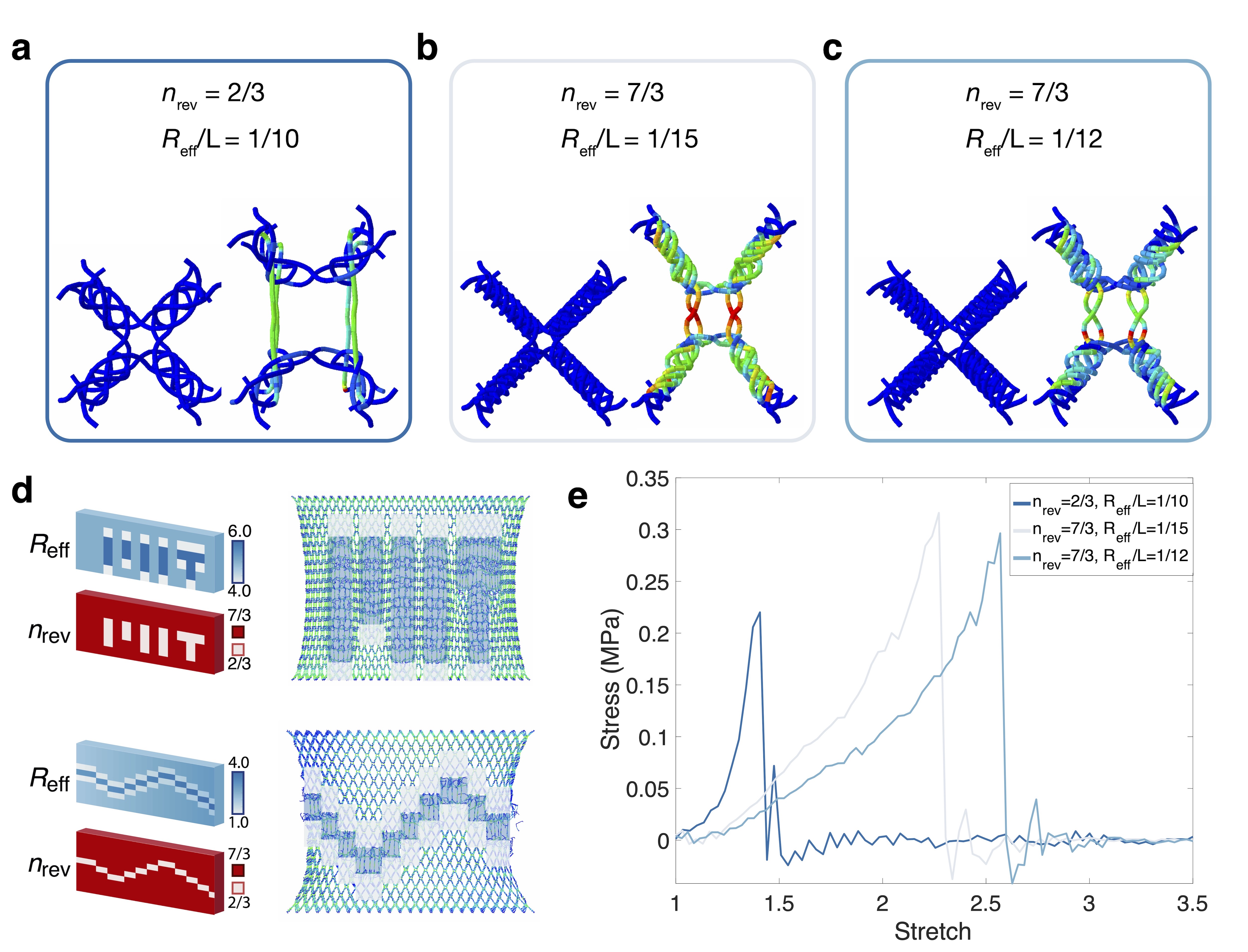}
\caption{\textbf{Constituent unit cells in Fig.~5.} \textbf{a--c}, The three unit cells composed to create Fig.~5a. 
Fig.~5b comprises the same unit cells, but with an additional gradient reducing the radius by a linear function, from -4 at the left boundary to -2 at the right boundary, in order to induce propagation of failure from left to right (i.e., $R_\mathrm{eff}$ spans from 1.0 to 4.0). Values of $n_\mathrm{rev}$ are unchanged. 
\textbf{d}, Arrangement of constituent unit cells in Fig.~5a-b. 
\textbf{e}, Stress-stretch curves of constituent unit cells in the form represented in Fig.~5a. The unit cell shown in (a) reaches its high-deformation regime (and hence failure) at the lowest stretch, about 1.4. This cell creates both a visible pattern (Fig.~5a) and a weakened region where failure is localized (Fig.~5b). The unit cells (b) and (c), with $n_\mathrm{rev}= 7/3$, have a larger stretch to failure, with (b), $R_\mathrm{eff} = 4$, slightly the stiffer of the two. This change in stiffness is used to create a boundary zone that sharpens the difference between cells (a) and (c).}
\label{fig:MITdemoUCs}
\end{figure}

\newpage
\subsection{References} 
\hfill \\

\newpage
\subsection{Supplementary videos \& captions}
\hfill

\noindent\textbf{Supplementary video 1.} \\
Comparison of beam-element simulation against {\it insitu} experimental tension of $2\times2\times2$ BCC woven lattice.

\noindent\textbf{Supplementary video 2.} \\
Evolution of curvature and contact pressure in beam-element simulations of $4 \times4 \times 4$ BCC woven lattice.

\noindent\textbf{Supplementary video 3.} \\
Beam-element simulation of ``MIT " patterned sample demonstrating architected deformation.

\noindent\textbf{Supplementary video 4.} \\
Beam-element simulation of patterned sample demonstrating architected lattice failure.

\noindent\textbf{Supplementary software.} \\
The design tool consists of two Python files. The top-level file, \url{lattice_main.py}, contains user-set parameters and options. The file
\url{makeunitcell_func1.py} contains functions required to construct woven lattices and output centerline data and 3D model files. 
\url{https://github.com/mollyacarton/woven-lattice}

\end{document}